\documentstyle[aaspp4,11pt]{article}
\newcommand{\beq}{\begin{equation}}
\newcommand{\eeq}{\end{equation}}
\newcommand{\beqa}{\begin{eqnarray}}
\newcommand{\eeqa}{\end{eqnarray}}

\newcommand{\Sub}[2]{\mbox{$#1_{\mbox{\scriptsize #2}}$}}
\newcommand{\lp}{\left(}
\newcommand{\rp}{\right)}
\newcommand{\etal}{{et al.}}
\newcommand{\ie}{{i.e.}}
\newcommand{\eg}{{e.g.}}
\newcommand{\cL}{\mbox{$\cal L$}}
\newcommand{\himpc}{\mbox{$h^{-1}$ Mpc}}
\newcommand{\Th}{T_{\rm he}}

\begin{document}
\title{Cosmological Simulations with Scale-Free Initial Conditions II:
Radiative Cooling}
\author{J. Michael Owen\altaffilmark{1}}
\affil{LLNL, L-38, P.O. Box 808, Livermore, CA 94551 \\
Email:  mikeowen@llnl.gov}
\altaffiltext{1}{Previous Address:  Ohio State University, Department of
Astronomy, Columbus, OH 43210}
\author{David H. Weinberg}
\affil{Ohio State University, Department of Astronomy, Columbus, OH 43210 \\
Email:  dhw@astronomy.ohio-state.edu}
\author{Jens V. Villumsen}
\affil{Max Planck Institut f\"{u}r Astrophysik, Karl Schwarzschild Strasse 1,
85740 Garching bei Munchen, Germany \\
Email:  victoria@infinet.com}

\begin{abstract}
The growth of structure from scale-free initial conditions is one of the
most important tests of cosmological simulation methods, providing a
realistically complex problem in which numerical results can be compared to
rigorous analytic scaling laws.  Previous studies of this problem have
incorporated gravitational dynamics and adiabatic gas dynamics, but
radiative cooling, an essential element of the physics of galaxy formation,
normally introduces a preferred timescale and therefore violates the
conditions necessary for self-similar evolution.  We show that for any
specified value of the initial power spectrum index $n$ [where $P(k)
\propto k^n$], there is a family of power-law cooling functions that
preserves self-similarity by ensuring that the cooling time in an object of
the characteristic mass $M_*$ is a fixed fraction $\hat{t}_C$ of the Hubble
time.  We perform hydrodynamic numerical simulations with an Einstein-de
Sitter cosmology, a baryon fraction of $5\%$, Gaussian initial conditions,
two different power spectrum indices, and four values of $\hat{t}_C$ for
each index, ranging from no cooling to strong cooling.  We restrict the
numerical simulations to two dimensions in order to allow exploration of a
wide parameter space with adequate dynamic range.  In all cases, the
simulations are remarkably successful at reproducing the analytically
predicted scalings of the mass function of dissipated objects and the gas
temperature distributions and cooled gas fractions in collapsed systems.
While similar success with 3-D simulations must still be demonstrated, our
results have encouraging implications for numerical studies of galaxy
formation, indicating that simulations with resolution comparable to that
in many current studies can accurately follow the collapse and dissipation
of baryons into the dense, cold systems where star formation is likely to
occur.
\end{abstract}


\section{Introduction}
Studies of evolution from scale-free initial conditions provide idealized 
but illuminating examples of the more general 
process of hierarchical structure
formation.  So long as the background cosmology, input physics, and initial
conditions are scale-free, even an inherently complex and highly nonlinear
process such as gravitationally driven, hierarchical structure formation
must evolve self-similarly in time.  Self-similar scaling offers a powerful
analytic guide to the behavior of
such complex systems, and investigations of
scale-free clustering have yielded important insights
concerning the growth of cosmological structure
(e.g., Davis \& Peebles 1977; Kaiser 1986; Efstathiou et al.\ 1988).
The evolution of scale-free initial conditions is also one of the
few cosmological problems in which numerical simulations can be tested
against rigorous analytic predictions, and the study of self-similar
gravitational clustering by Efstathiou et al.\ (1988) is one of the
key pieces of evidence for the accuracy of cosmological N-body methods.

In Owen
\etal\ (1998b; hereafter Paper I), we extended this approach to adiabatic
gas dynamics, presenting a set of smoothed particle hydrodynamics (SPH)
simulations designed to study self-similar evolution of structure in a mixed
baryon/dark matter universe.  We found that the resulting population of
collapsed structures demonstrated the expected self-similar
scalings, so long as we were careful to account for the numerical
limitations of each experiment.  However, while the models presented in
Paper I included gravitational, pressure, and shock processes, they
neglected the effects of radiative energy loss from the gas, 
a critical element in the formation and evolution of galaxies (White \&
Rees 1978).  In this paper we extend our earlier work to include 
radiative dissipation; for each choice of the scale-free initial 
power spectrum $P(k) \propto k^n$, we construct artificial cooling
laws that maintain the scale-free nature of the physics.

Our approach in this investigation differs from that of Paper I in a
few important respects.  First, in Paper I we considered a set of 3-D
simulations, while for this study we restrict ourselves to 2-D simulations.
The restriction to 2-D allows us to perform a larger number of 
high dynamic range experiments than would be practical in 3-D.  
While 3-D simulations are clearly necessary for realistic studies
of galaxy formation, for our present purposes we wish to study
the effects of radiative dissipation in a variety of
idealized, scale-free models, and 2-D
experiments provide an economical starting point.  
Our investigation represents a first attempt to examine self-similar
evolution in models that incorporate the physical processes most
essential to galaxy formation: gravitational collapse and merging,
shock heating, and radiative cooling.  We will use our results to
guide the choice of parameters for more expensive, 3-D models in a 
future study.

Section 2 discusses the analytic scaling of characteristic group properties
in 2-D models.  In Section 3 we derive the cooling functions that preserve
self-similarity, for both 2-D and 3-D models.  Section 4 describes the
simulations and their numerical limitations, and it presents our first
important numerical result, the scaling of the mass functions of dark 
matter, baryon, and dissipated baryon groups.  Section 5 presents tests
of mass, temperature, density, and Bremsstrahlung luminosity scaling, 
similar to those used in Paper I.  Section 6 examines the central issue
of the paper, the scaling of the cooled baryon component in collapsed
objects.  In Section 7 we summarize our conclusions and discuss their
implications for numerical studies of galaxy formation.  

\section{Expected Scalings for 2-D Perturbations}
\label{2dscale.sec}
While working in 2-D affords us a number of computational advantages,
studying cosmological structure in 2-D introduces a number of subtle
departures from the more familiar 3-D case.
It is important to understand that these models really
represent the imposition of 2-D perturbations 
in a formally 3-D universe.  
The correct conceptual way to view the setup for these simulations
is that we are taking an initially homogeneous, 3-D universe containing 
baryons and dark matter and initializing only $k_x$ and $k_y$ density
perturbations in Fourier space.  
These initial conditions imply that structure evolves in the 
$x$ and $y$ directions alone -- matter continues to simply expand with the
Hubble flow in the $z$ direction.  Therefore, the most collapsed structures
that form represent infinitely long filaments, rather than collapsed
groups or galaxies.  When we measure the ``mass'' of one of these 2-D
objects, we are in fact measuring a mass per unit length, or linear density
$\theta \propto m/\ell$.  Similarly, the mass of each particle in the
simulations is actually a mass per unit length, and the particles interact
gravitationally as though they were infinite, thin rods.

To illustrate some of the more important distinctions from the general 3-D
case, consider the collapse and formation of an isolated filament.  If the
total mass per unit length of this object is $\theta$, and we have another
infinite, thin rod of mass per unit length $\Theta$ at a distance $r$, then
their mutual gravitational force is $F = -G \theta \Theta/r \propto 1/r$,
and the gravitational potential of the first rod
is $\Phi(r) = -G \theta \ln r$.  The
circular velocity, $v_c^2 = r d\Phi/dr = G \theta$, is independent of
distance.  In order to define a temperature that corresponds to this
circular velocity (see Thoul \& Weinberg 1996 for the 3-D analogy),
consider an isothermal atmosphere of gas about the filament in hydrostatic
equilibrium.  If the gas follows a density profile 
$\rho(r) = \rho_1 (r/r_1)^{-\alpha}$
and the pressure is given by $p(r) = k_B T (\mu m_p)^{-1} \rho(r)$, then the
requirement of hydrostatic equilibrium becomes
\beq
  \frac{G \theta}{r} 2 \pi r ~dr~ \rho(r) = -\frac{dp}{dr} 2 \pi r ~dr,
\eeq
which after some manipulation yields
\beq
  \label{Tvc2d.eq}
  T ~=~ \frac{\mu m_p}{\alpha k_B} G \theta ~=~ \frac{\mu m_p}{\alpha k_B}v_c^2.
\eeq
Inspection reveals that typically our 2-D objects obey $\rho(r) \propto
r^{-2}$ (where $\rho$ is the mass per unit {\em volume}), so we recover
the 3-D result, $T = \mu m_p (2 k_B)^{-1} v_c^2$.

If the input physics, background cosmology, and initial conditions
are scale-free, then at a given time the only characteristic physical
scale is set by the amplitude of density fluctuations.
This condition implies scaling laws for the time evolution of radii,
masses, temperatures, and Bremsstrahlung luminosities of collapsed objects.
We derived these scaling laws in Paper I, following the reasoning
of Kaiser (1986), but here we must reformulate them for 2-D perturbations.
The variance of (linear theory)
mass fluctuations smoothed with a window of comoving scale $R^c$ is
\beq
  \label{sigma.eq}
  \sigma^2_M(R^c,a) = \int_0^\infty 2\pi k dk P(k,a) W(kR^c) \propto
  a^2 (R^c)^{-(n+2)},
\eeq
where $P(k,a) \propto a^2 k^n$ is the linear fluctuation power spectrum.
We can define a characteristic comoving length scale $R_*^c$ to
be the scale on which the linear fluctuation variance has some specified value.
Equation~(\ref{sigma.eq}) implies that $R_*^c \propto a^{2/(2+n)}$,
and the corresponding radius in physical units is
\beq
  \label{Rscale2d.eq}
  R_* \propto a R_*^c \propto a^{(4 + n)/(2 + n)}.
\eeq
Since the underlying
physics is 3-D, the proper background density scales as $a^{-3}$, and we
have
\beq
  \label{rhoscale2d.eq}
  \rho_* \propto \bar{\rho} \propto a^{-3}.
\eeq
The characteristic linear density $\theta_*$ (analogous to the
characteristic mass $M_*$ in 3-D) scales as
\beq
  \label{Mscale2d.eq}
  \theta_*^c \propto \rho_*^c (R_*^c)^2 \propto a^{4/(2 + n)} 
  ~~\Longrightarrow~~
  \theta_* \propto a^{-1} \theta_*^c \propto a^{(2 - n)/(2 + n)}.
\eeq
The difference between the physical $\theta_*$ and the comoving $\theta_*^c$
is the $a^{-1}$ factor, which reflects the drop of linear mass densities
caused by expansion along the 3rd axis.
Based on equation (\ref{Tvc2d.eq}), we know that the hydrostatic
equilibrium temperature scales directly with $\theta$, so
\beq
  \label{Tscale2d.eq}
  T_* \propto \theta_* \propto a^{(2 - n)/(2 + n)}.
\eeq
Finally, the Bremsstrahlung luminosity per unit length $\cL \propto
L/\ell$ scales as
\beq
  \label{Lscale2d.eq}
  \cL \propto \theta_* \rho_* T_*^{1/2} 
  \propto a^{-(6 + 9n)/(4 + 2n)}.
\eeq
Note that we are defining the luminosity to be proportional to the square-root
of the temperature.  Even though we will be assuming a different
temperature dependence for radiative cooling in the following section,
whenever we refer to the ``luminosity'' we will still use $\cL \propto
T^{1/2}$, since this corresponds to the physical process of Bremsstrahlung
emission in the real universe.

The $*$-ed quantities in the proportionality 
relations~(\ref{Rscale2d.eq})-(\ref{Lscale2d.eq}) could refer to any
specific choice of the linear fluctuation variance.  In \S 4 
we will give a precise definition of $\theta_*$ in terms of the
characteristic mass appearing in the Press-Schechter (1974) mass function
formula (eq.~[\ref{fPS2d.eq}] below).  We will also refer frequently
to the mass scale $\Sub{\theta}{nl}$ on which the fluctuation variance
is unity, indicating that rms fluctuations on this scale are entering
the non-linear regime.  Specifically, this mass scale is defined by
the conditions
\beq
  \label{ThetanlDef.eq}
  \Sub{\theta}{nl}(a) = \pi R^2_{\rm nl} \bar{\rho}(a),
\eeq
where
\beq
  \label{RnlDef.eq}
  \sigma^2(\Sub{R}{nl},a)=1.
\eeq

\section{Radiative Cooling Laws}
\label{CoolLaw.sec}
A realistic cooling function like the one arising in a
primordial H/He plasma would
impose a dimensional physical scale into our calculations, violating the
conditions that lead to self-similar time evolution.  
The clearest demonstration of this problem is that the cooling
time in an object of characteristic mass $M_*$ would not be a constant fraction 
of the Hubble time, since the standard cooling function has a complicated
temperature dependence and the characteristic temperature $T_*$ and
density $\rho_*$ evolve in time.  For example, at some epochs most of
the baryonic mass in an $M_*$ object might be able to cool in a Hubble
time, while at other epochs only a small fraction would be able to cool.
We therefore cannot
use the standard physical cooling law and maintain the scale-free nature of
the problem.

Nonetheless, for a given power-law initial power spectrum,
$P(k) \propto k^n$, it is possible to determine a power-law cooling
relation $\Lambda(T)/\Sub{n}{H}^2 \propto T^\beta$ such that the cooling
time in an $M_*$ object {\it is}
a fixed fraction of the Hubble time.  So long as the cooling law meets
this requirement, the physical system remains scale-free, and the
rigorous prediction of self-similar evolution for the system holds.  Under
this condition, at a fixed expansion factor the ratio of the cooling time
to the Hubble time may vary with mass, but the ratio depends only
on $M/M_*$ and thus scales properly with time.  
The dependence of the cooling exponent $\beta$ on the power spectrum
index $n$ is different for 2-D and 3-D perturbations, so we will give 
the results for both cases here.

The radiative contribution to the time rate of change of the specific
thermal energy $u$ is
\beq
  \left.\frac{Du}{Dt}\right|_R = \frac{\Lambda(T)}{\rho},
\eeq
where we use the subscript $R$ to denote the fact that this is only the
radiative piece of $Du/Dt$.  We parameterize the radiative cooling
function $\Lambda(T)$ as a power-law,
\beq
  \frac{\Lambda(T)}{\Sub{n}{H}^2} 
    = \left[\mu m_p (1+\Sub{n}{He}/\Sub{n}{H}) \right]^2 
      \frac{\Lambda(T)}{\rho^2}
    \equiv A_0 T^\beta
    = A_1 u^\beta.
\eeq
In these terms the cooling time $t_C$ can be expressed as
\beq
  t_C = u \lp \left.\frac{Du}{Dt}\right|_R \rp^{-1}
      = u \lp \frac{\Lambda(T)}{\rho} \rp^{-1}
      = u (\rho A_1 u^\beta)^{-1}
      = A_1^{-1} \rho^{-1} u^{1 - \beta}.
\eeq
The requirement to maintain self-similarity is that the cooling time 
$t_C^*$ for a
characteristic $\theta_*$ (or $M_*$) object must be a fixed fraction
$\hat{t}_C$ of the Hubble time $t_H = H_0^{-1} (a/a_0)^{3/2}$.  Note that in
general we expect the cooling time to be a function of mass --- e.g., 
$\hat{t}_C(0.5 ~\theta_*) \ne \hat{t}_C(2 ~\theta_*)$ --- but we are forcing
the cooling time for any given multiple of a characteristic mass to be
fixed at all times, so that
$\hat{t}_C(2 ~\theta_*, a_1) = \hat{t}_C(2 ~\theta_*, a_2)$.  

For 2-D perturbations we know that the specific thermal energy and
the density scale as $u_* \propto a^{(2-n)/(2+n)}$ and $\rho_* \propto
a^{-3}$, so fixing $t_C^* = \hat{t}_C t_H$ yields,
after some manipulation, 
\beq
  \label{f2d.eq}
  A_1^{-1} \rho_0^{-1} u_0^{1 - \beta}
  = \hat{t}_C H_0^{-1} \lp\frac{a}{a_0}\rp^{-3/2 - (1 - \beta)(2 - n)/(2 + n)}.
\eeq
Our requirement that $t_C^*$ remain the same fraction of the Hubble time
at all times implies that the $a$ dependence on the right-hand side of
equation (\ref{f2d.eq}) must vanish, which finally gives us
\beq
  \label{beta2d.eq}
  \Sub{\beta}{2-D} = \frac{3}{2} ~\frac{2 + n}{2 - n} + 1.
\eeq
The 3-D derivation is very similar, except that for 3-D perturbations the
specific thermal energy scales as $u_* \propto a^{(1 - n)/(3 + n)}$.
Following the same arguments as above yields
\beq
  \Sub{\beta}{3-D} = \frac{3}{2} ~\frac{3 + n}{1 - n} + 1.
\eeq

Now that we know the appropriate index for the cooling power-law, we need
only specify the normalization.  According to equation (\ref{f2d.eq}), the
normalization $A_1$ is proportional to
\beq
  \label{Anorm.eq}
  A_1 \propto \frac{H_0}{\hat{t}_C \rho_0 u_0^{\beta - 1}} 
      \propto \frac{H_0}{\hat{t}_C \rho_0 T_0^{\beta - 1}},
\eeq
where $\rho_0$ and $T_0$ are chosen as characteristic of a fiducial group
at expansion $a=a_0$.  We will adopt the convention that the characteristic
density is $\rho_0 \sim 1000 \Sub{\bar{\rho}}{bary}$, as this is
typical of the objects we select.  A well-defined choice for the
characteristic temperature is the hydrostatic equilibrium
support temperature for an object with
the nonlinear mass, 
\beq
  \label{TnlDef.eq}
  \Sub{T}{nl} ~\equiv~  \Th(\Sub{\theta}{nl}) ~= ~
  \mu m_p (2 k_B)^{-1} G \Sub{\theta}{nl}.
\eeq
Then by choosing a cooling time fraction $\hat{t}_C \equiv t_C/t_H$, we
completely specify the normalization of the cooling law.  
For $\hat{t}_C \ll 1$, most of the baryon mass in $\Sub{\theta}{nl}$
objects should be able to cool, while for $\hat{t}_c \gg 1$ most of the
baryons in a $\Sub{\theta}{nl}$ object will be unable to cool before the
it merges with another of comparable mass, shock heating the gas
to a higher temperature.

\section{The Simulations}
\label{Sims.sec}
We examine a number of different physical scenarios, based upon
two distinct sets of initial density perturbations and various choices for
the amplitude of the associated cooling law.  As in Paper I, the background
cosmology is a flat, Einstein-de Sitter universe with
$\Sub{\Omega}{bary}=0.05$ and $\Sub{\Omega}{dm}=0.95$.  We consider two
sets of initial density perturbations, each Gaussian distributed, with
power-law power spectra of $n=0$ and $n=+1$ [where $P(k) \propto k^n$].  
Dynamically these correspond to 3-D perturbation spectra of indices
$\Sub{n}{3-D} = -1$ and 0, respectively, for quantities represented by 
integrals over the power spectrum (such as the rate at which the scale of
nonlinearity grows with time).  

The initial amplitude of the
density fluctuations is chosen so that the linearly extrapolated mass
fluctuations at the end of the simulation
have an rms amplitude of unity for a top-hat filter of
radius $\Sub{R}{th} \approx 0.1 \Sub{L}{box}$ for $n=0$ and $\Sub{R}{th}
\approx 0.3 \Sub{L}{box}$ for $n=+1$, where
$\Sub{L}{box}$ is the simulation box size.  Throughout this paper we
parameterize the evolution in terms of the expansion factor $a$, defined so
that the final output expansion in each scenario is $a_f \equiv 1$.  In
principle, because of the scale-free nature of these experiments, there are
many possible choices for identifying the simulation box size with a
physical scale in the real universe.  For instance, if we identify the $a_f
\equiv 1$ output with the observed universe at $z=0$ and adopt the
normalization condition $\sigma_8 = 0.5$ (appropriate for the rms mass
fluctuation in spheres of radius 8 \himpc\ with $\Omega = 1$; see White,
Efstathiou, \& Frenk 1993), then the implied comoving box size is $\Sub{L}{box}
\approx 40 ~\himpc$ for the $n=0$ models, and $\Sub{L}{box} \approx 17
~\himpc$ for $n=+1$.  Earlier output times can be identified with higher
redshifts, $z = a^{-1} - 1$.  Alternatively, one can identify any expansion
$a$ with the $z=0$ universe, in which case the implied box size (for
$\sigma_8 = 0.5$) is $\Sub{L}{box} = 40 ~a^{-1} ~\himpc$ for $n=0$, and
$\Sub{L}{box} = 17 ~a^{-2/3} ~\himpc$ for $n=+1$, decreasing steadily as the
nonlinear scale becomes a larger fraction of the box size.

For both the $n=0$ and $n=+1$ initial conditions, we perform simulations
with several amplitudes of the cooling law, parameterized in terms of the
dimensionless cooling time $\hat{t}_C = t_C/t_H$ as
$\hat{t}_C = \infty$ (\ie, no cooling), 200, 1, and 0.1.  All simulations
are performed using $\Sub{N}{bary} = \Sub{N}{dm} = 128^2$ computational
nodes, except for one case where we repeat the $n=0$, $\hat{t}_C = 1$ model
with $\Sub{N}{bary} = \Sub{N}{dm} = 256^2$ nodes in order to check for
convergence.  The gravitational interactions are evaluated on a $512^2$
Particle-Mesh (PM) grid, and the hydrodynamical interactions are modeled
using Adaptive Smoothed Particle Hydrodynamics (ASPH).  For a complete
description of the code and techniques used, see Owen \etal\ (1998a).  In
all we present nine simulations of eight distinct physical models (2 sets
of initial density perturbations $\times$ 4 possible cooling law
amplitudes).

We examine each simulation at expansions spaced logarithmically, roughly
with an interval of $\Delta \log a = 0.2$.  We are primarily interested in
the properties of the collapsed objects that form in each model,
particularly in those that undergo significant radiative dissipation.
Therefore, as in Paper I, at each expansion we identify groups of particles
using the ``friends-of-friends'' algorithm (see, \eg, Barnes \etal\ 1985;
the specific implementation used here can be found at
{\bf http://www-hpcc.astro.washington.edu/tools/FOF/}).  We compute global
average properties for each group (such as the dark matter and baryon mass,
the mass and emission weighted temperatures, luminosity, and so forth) and
check for self-similar evolution in the distributions of properties of
these groups.  Although there are more sophisticated group finding algorithms
available, friends-of-friends provides a simple, flexible, and unambiguous
definition of a group using an algorithm that maintains the conditions
necessary for self-similar scaling (because it does not introduce a fixed
physical scale).  In analyzing these experiments we use linking parameters
of $\ell = 0.2 ~\Delta x_p$ (the same as used in Paper I) and $\ell = 0.05
~\Delta x_p$.  Groups selected with $\ell = 0.2 ~\Delta x_p$ consist of
both hot and dissipated gas, whereas $\ell = 0.05 ~\Delta x_p$ tends to
select objects that have undergone significant radiative dissipation.
There are essentially no dark matter objects found
with $\ell = 0.05 ~\Delta x_p$, and very little baryonic mass
collapses to this density in the models without cooling.

\begin{figure}[htbp]
\plottwo{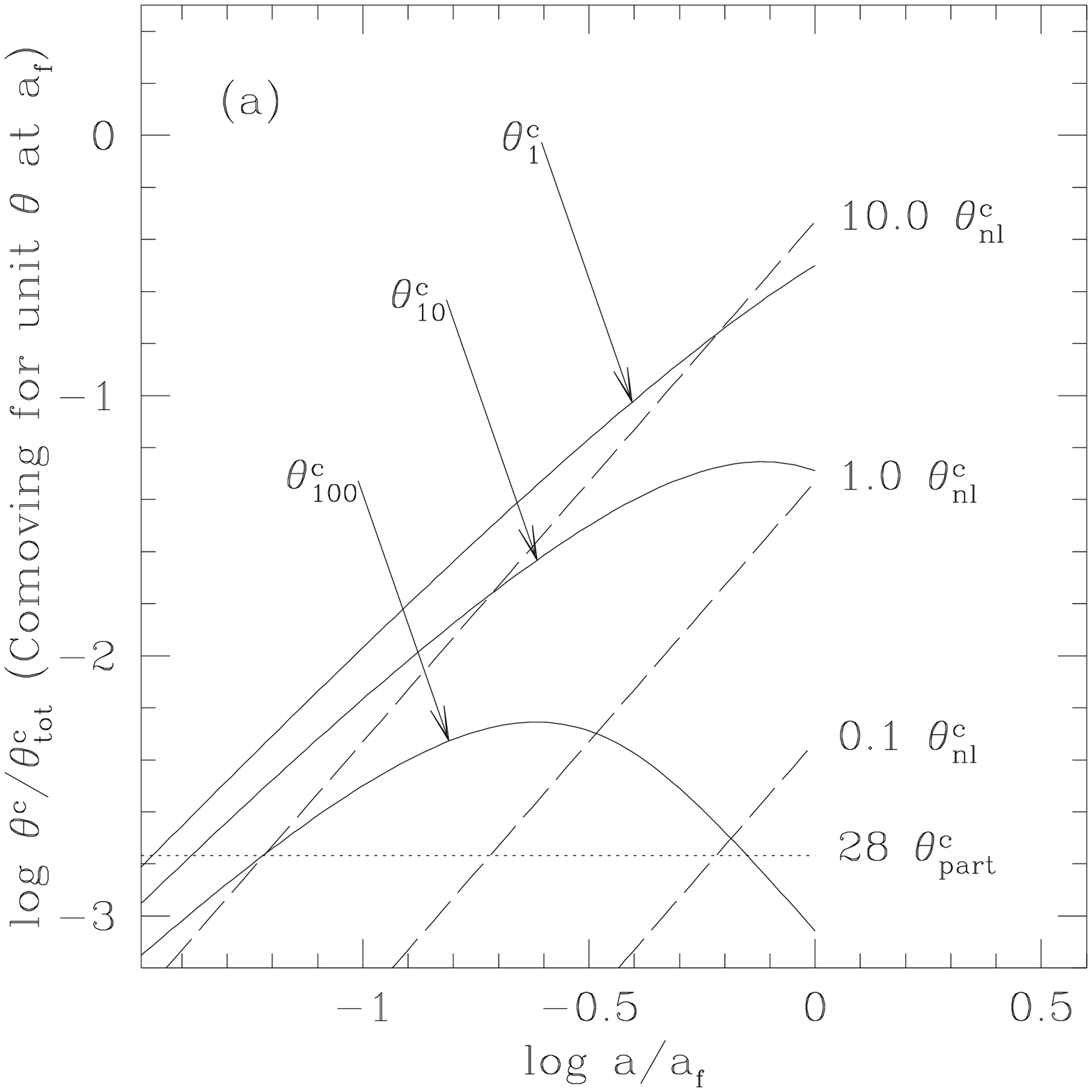}{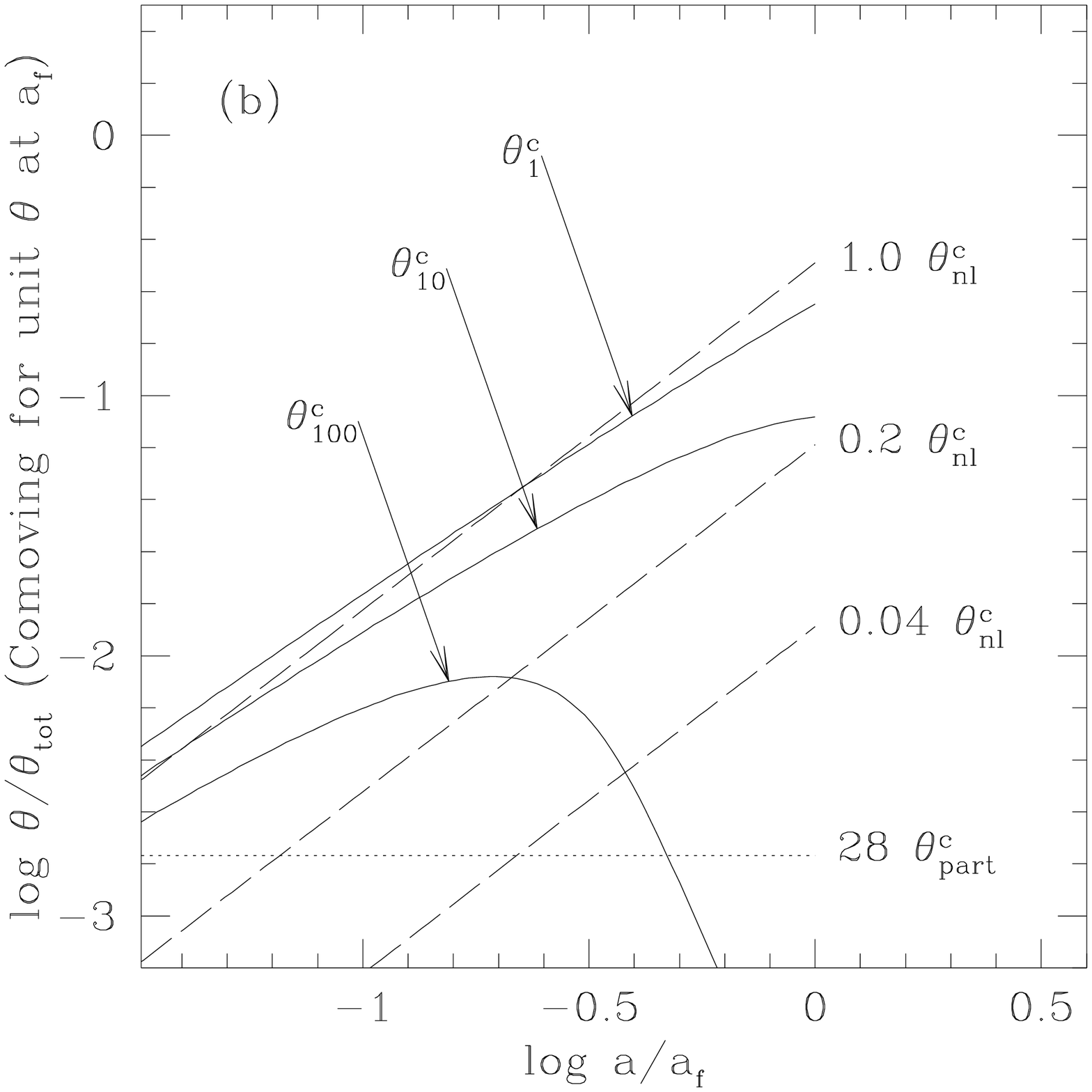}
\caption{Comoving $\theta$ resolution limits as a function of expansion for
the (a) $n=0$ and (b) $n=+1$ simulations.  Although 
$\theta \propto m/l$ is a linear mass density, in this 2-D context
it plays the same
role as the mass in 3-D.  The horizontal dotted lines show the low $\theta$
limit, assumed to be 28 times the particle ``mass'' for the
$\protect\Sub{N}{bary} = \protect\Sub{N}{dm} = 128^2$ simulations (actually
the linear mass density per particle: $\theta_i = \protect\Sub{L^2}{box}
~\bar{\rho}/N$).  The dashed lines show various multiples of the nonlinear
mass scale $\protect\Sub{\theta}{nl}$, while the solid lines show the PS
prediction such that statistically we would expect to find 1 ($\theta_1$),
10 ($\theta_{10}$), and 100 ($\theta_{100}$) objects of that mass in the
simulation volume.  All $\theta$'s are scaled such that the total $\theta$
in the volume is 1.}
\label{MassRes.fig}
\end{figure}
In order to understand the regimes we can probe, we must first identify the
mass resolution limits of our experiments.  Since ASPH is a Lagrangian
technique, the lower limit on the hydrodynamic interactions is best
expressed as a mass limit, set by a multiple of the particle mass.  In each
experiment the ASPH smoothing tensors are evolved so that each node sees
significant contributions from roughly 28 of its neighbors, which provides
a reasonable lower limit on the ASPH mass resolution.  The gravitational
force resolution is effectively determined by a multiple of the
gravitational softening length, in this case the size of a PM grid cell, so
unfortunately this is not Lagrangian in the same way as the hydrodynamic
resolution.  However, the PM grid has a linear resolution of 512 elements
across the system, and on the scale of the objects we will be examining we
may consider the gravitational resolution to be unrestrictive.  As a check
of this assumption, we have experimentally verified that the results are
insensitive to changes in the gravitational PM resolution by factors of two
(a factor of 4 in the total number of PM cells).

The upper limit on the mass range we are sensitive to is set by our box
size.  The larger a group is, the more statistically rare it is, and the
less likely we are to find examples of such structures in any finite,
randomly realized volume.  We can therefore only expect to find objects up
to a certain size at any given output time within our finite simulation
volume.  The Press-Schechter mass function (Press \& Schechter 1974;
hereafter PS) provides a rough estimate of this limit.  
However, there is a subtle
distinction between using the 2-D version of PS theory (eq.~[\ref{fPS2d.eq}]
below) and using the familiar, 3-D PS formalism.  In 3-D,
friends-of-friends with a linking parameter $\ell = 0.2 \Delta x_p$ selects
objects within an overdensity contour of roughly $\delta \equiv \delta
\rho/\rho \sim 250$.  The well-known solution for the collapse of an
isolated 3-D perturbation (the ``top-hat'' collapse model) tells us that an
object with an actual overdensity of $\delta \sim 250$ corresponds to a
region in the initial conditions with a 
linearly extrapolated mass fluctuation of $\delta_* = 1.68$.  We can therefore
use $\delta_* = 1.68$ in the PS mass distribution to predict the properties
of the population of objects identified by friends-of-friends.  In 2-D,
however, the familiar top-hat solution does not apply, so we do not have an
{\em a priori} prediction for the appropriate value of $\delta_*$ to plug
into the PS mass function.  Instead, we measure the differential group mass
distribution $f(\theta)$ in the simulations and fit the PS model by varying
the value of $\delta_*$, the results of which can be seen in Figure
\ref{fM.fig} below.  We find that values of $\delta_* \approx 0.7$ for
$n=0$ and $\delta_* \approx 3$ for $n=+1$, are appropriate.  

With this
knowledge, we can plot a set of lower and upper comoving $\theta$ limits
for each set of initial conditions, as shown in Figure \ref{MassRes.fig}.
This figure also shows the evolution of various multiples of the comoving
nonlinear ``mass'' $\Sub{\theta^c}{nl}$, defined earlier in
equation~(\ref{ThetanlDef.eq}).
Based on these results, for $n=0$
we can expect most of the mass range $\theta^c \in [1 ~\Sub{\theta}{nl},
10 ~\Sub{\theta}{nl}]$ to be accessible over a range of expansions $\log
a/a_f \in [-0.7,-0.1]$.  Similarly, for $n=+1$ we find $\theta^c \in [0.1
~\Sub{\theta}{nl}, 1 ~\Sub{\theta}{nl}]$ should be accessible over $\log
a/a_f \in [-1,0]$.

\begin{figure}[htbp]
\plottwo{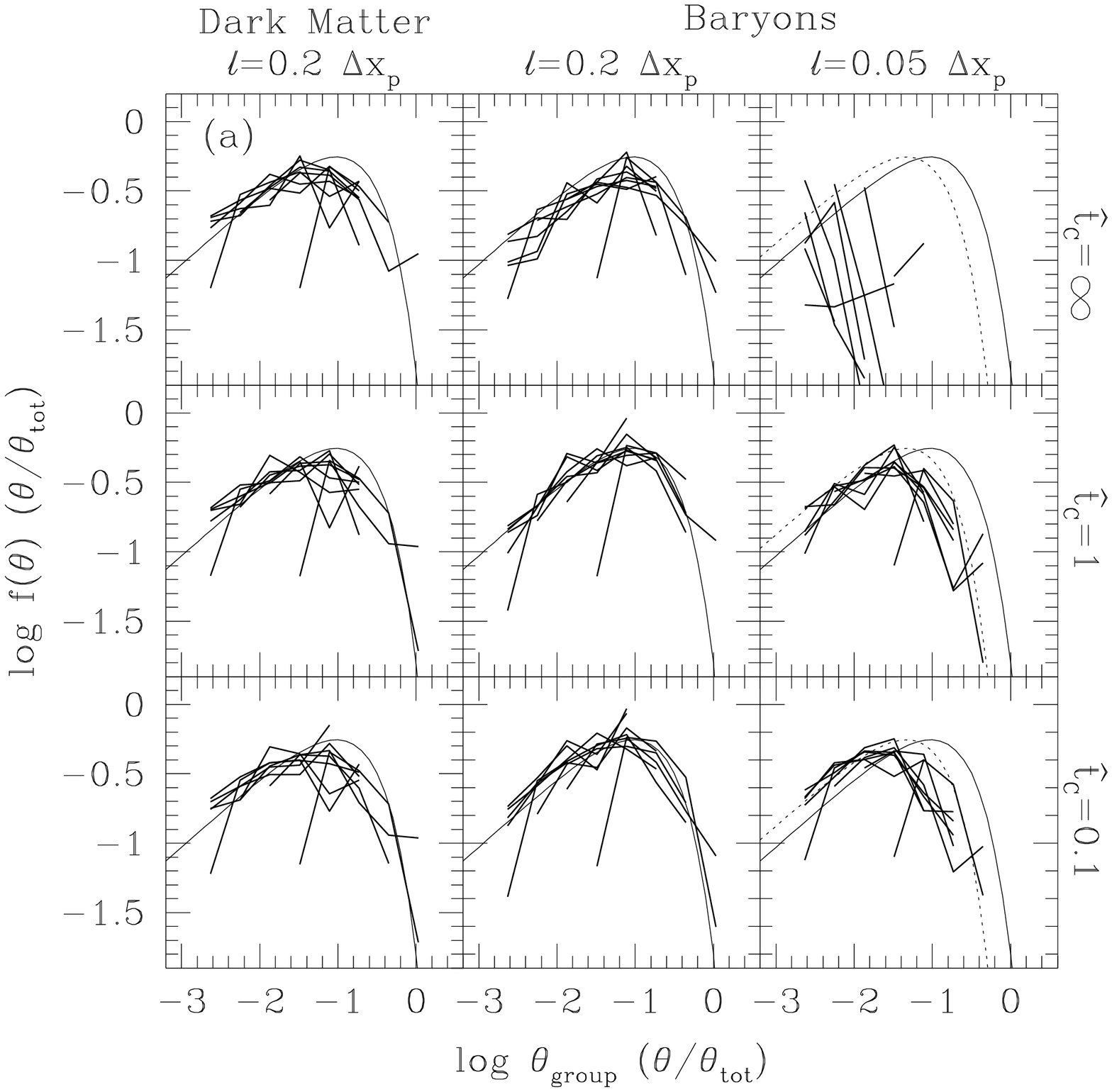}{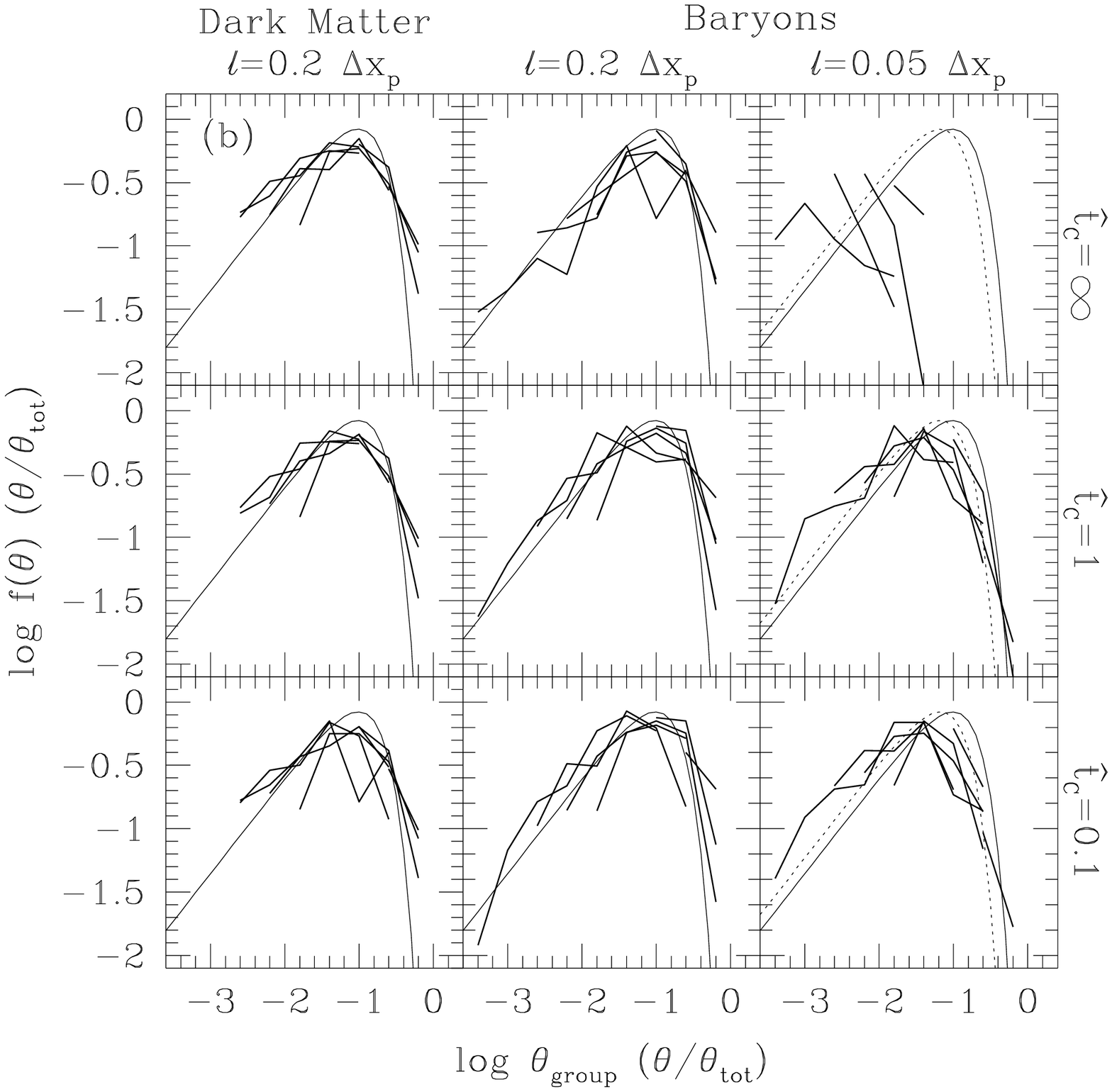}
\caption{Differential mass distribution function $f(\theta)$ for the (a)
$n=0$ and (b) $n=+1$ simulations, scaled to $a=a_f$.  In each figure the
left column shows the dark matter groups linked with $\ell = 0.2 ~\Delta
x_p$, the middle column the baryon groups linked with $\ell = 0.2 ~\Delta x_p$,
and the right column the baryon groups linked with $\ell = 0.05 ~\Delta
x_p$.  The heavy solid lines show the measured $f(\theta)$ at various
expansions, shifted (assuming self-similarity) to $a=a_f$.  The thin solid
lines show the PS mass function (eq.~[\ref{fPS2d.eq}]) assuming (a)
$\delta_* = 0.7$ and (b) $\delta_* = 3$, fitted as the appropriate
comparison for objects identified with linking parameter $\ell = 0.2 ~\Delta
x_p$ for the $n=0$ and $n=+1$ models,
respectively.  Similarly, the thin dotted lines in the $\ell = 0.05 ~\Delta
x_p$ panels show the PS mass function using (a) $\delta_* = 1$ ($n=0$) and
(b) $\delta_* = 4$ ($n=+1$).}
\label{fM.fig}
\end{figure}
In Figure \ref{fM.fig} we plot the mass distribution $f(\theta)$ of objects
identified by friends-of-friends
over a range of expansion factors for the simulations without cooling
($\hat{t}_C \equiv \infty$) and for the cooling models with
$\hat{t}_C=1$ and $\hat{t}_C=0.1$.  Note that $f(\theta)$ is the amount of mass
contained in groups in each mass range, not the number of objects, and that
we plot the baryon and dark matter groupings separately.  At each expansion
factor
$f(\theta)$ is calculated and shifted (assuming that the self-similar relation
$\theta \propto a^{(2-n)/(2+n)}$ holds) to $a=a_f$ for comparison, giving
the heavy solid lines.  If the simulations perfectly obeyed the
analytically predicted self-similar scalings, then the plotted mass functions
would connect together into a continuous relation.  There are no free
parameters and no approximations in this scaling, 
and, as with the adiabatic results in Paper I,
the measured distributions $f(\theta)$ at different expansions link up
nicely.  Therefore, it appears that the group mass distribution is following
the expected self-similar behavior well.  The only notable exception is the
$\ell = 0.05 ~\Delta x_p$ groups in the models without cooling (the upper
right panels).  This failure is to be expected, though, since very 
little mass in the non-dissipative models collapses to such high density, and
therefore the statistical sampling in this regime is poor.

The thin lines in Figure \ref{fM.fig} show the 2-D PS prediction, given by
the relation
\beq
  \label{fPS2d.eq}
  f(\theta) ~d\theta = \sqrt{\frac{2}{\pi}} \lp \frac{2 + n}{4} \rp \lp
    \frac{\theta}{\theta_*} \rp^{(2 + n)/4} \exp \lp -\frac{1}{2} \lp
    \frac{\theta}{\theta_*} \rp^{(2 + n)/2} \rp \frac{d\theta}{\theta},
\eeq
where $f(\theta) ~d\theta$ represents the mass fraction of groups in the
mass range $[\theta, \theta + d\theta]$.  Here $\theta_*$ is defined 
similarly to $\Sub{\theta}{nl}$ in equations~(\ref{ThetanlDef.eq})
and~(\ref{RnlDef.eq}), except that the variance in equation~(\ref{RnlDef.eq})
is set to $\delta_*^2$ instead of to unity.  Altering $\delta_*$
slides the resulting PS curve horizontally in this Figure but does
not change its shape.  Since, as mentioned above, we do not have
the 3-D spherical collapse model as a guide, we fit the PS prediction to the
measured $f(\theta)$ curves with $\delta_*$ as a free parameter,
obtaining $\delta_* = 0.7$ for $n=0$ and $\delta_* = 3$ for $n=+1$.
Equation~(\ref{sigma.eq}) shows that the relation between $\theta_*$
and our other fiducial ``mass'' $\Sub{\theta}{nl}$ is 
\beq
  \label{ThetastarDef.eq}
  \theta_* = \Sub{\theta}{nl} \delta_*^{4/(2+n)}.
\eeq

For the $\ell = 0.2 ~\Delta x_p$ objects, PS curves
with the same value of $\delta_*$ fit the dark matter and
baryonic objects equally well.  This simultaneous fit
implies that the baryon-to-dark matter
ratio in these structures is near the universal average, 
in contrast to the results of Paper I,
where we found that the baryons in groups tended to be underrepresented
relative to the dark matter.  
The difference might reflect either the 2-D geometry or
the somewhat higher resolution of our present experiments.
There is also a hint in Figure~\ref{fM.fig} (and in Figure~\ref{Mscale.fig}
below) that the baryon fractions are slightly higher in the models
with radiative cooling.

The thin dotted lines in the right columns of Figure~\ref{fM.fig}
show the PS prediction using $\delta_* = 1$ for $n=0$ and $\delta_* = 4$
for $n=+1$.  These dotted lines match the distribution of mass in
dissipated gas objects (as selected by $\ell = 0.05 ~\Delta x_p$)
reasonably well, despite the fact that there is no rigorous justification for
using PS theory to make predictions about such dissipative structures --- PS
theory is built on an approximate description of purely gravitational dynamics.
Since there is a free parameter in our PS fits and it is chosen separately
for $\ell=0.05\Delta x_p$ and $\ell = 0.2 \Delta x_p$, it is not clear
how much significance should be attached to this match.  It does appear,
however, that the relation between ``dissipated mass'' 
(identified by $\ell = 0.05 \Delta x_p$)
and ``virialized mass'' (identified by $\ell = 0.2 \Delta x_p$)
is simple enough in these scale-free models that PS theory gives
a reasonable guide to the mass function of the cooled objects.

\section{Testing for Self-Similar Evolution}
\begin{figure}[htbp]
\plottwo{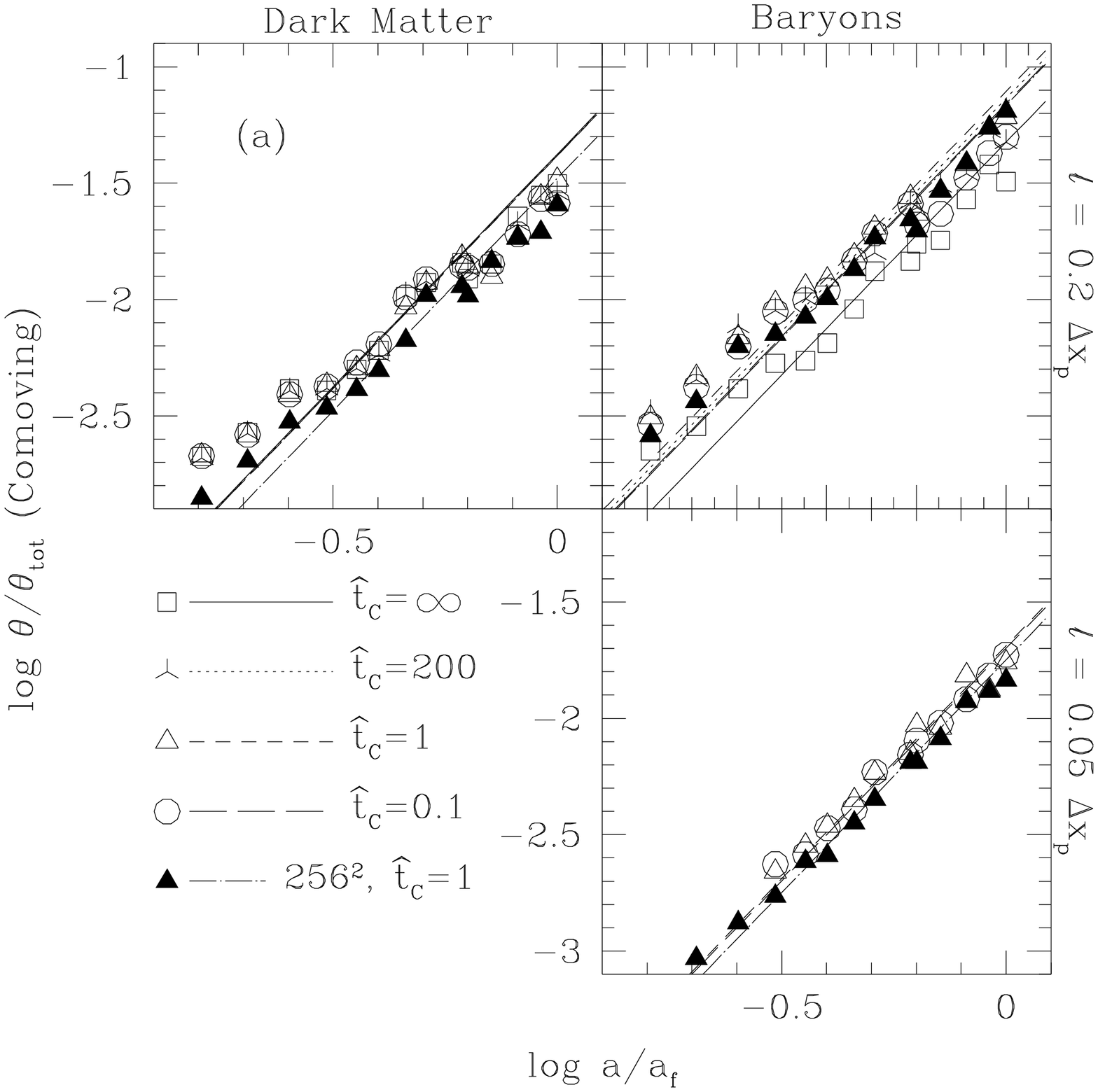}{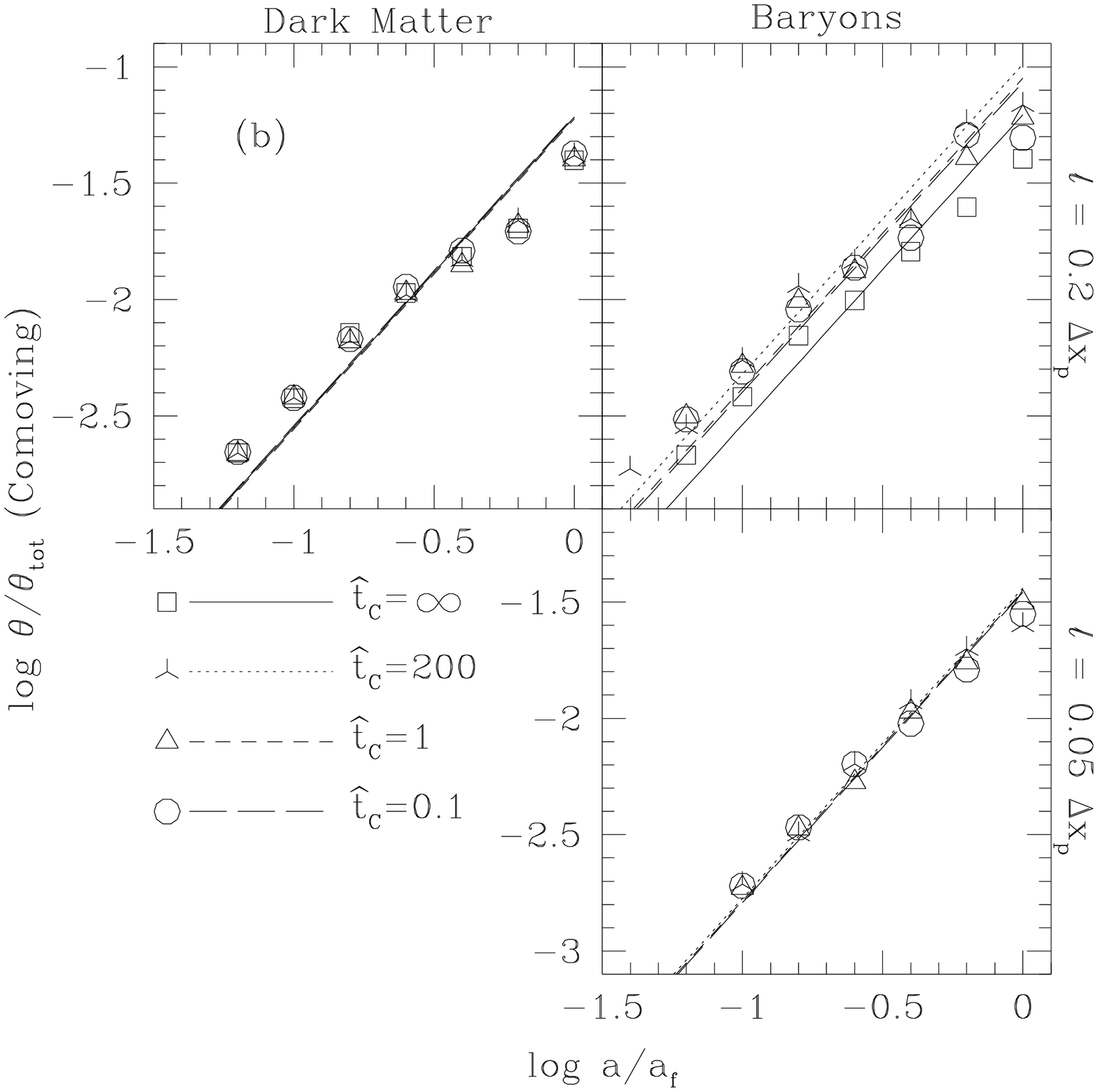}
\caption{Evolution of the fiducial comoving linear mass density
$\theta^c_{70\%}$ for the (a)
$n=0$ and (b) $n=+1$ models, where $\theta^c_{70\%}$ is defined so that at
each output time $70\%$ of the mass in the simulation is contained in groups
with $\theta^c \le \theta^c_{70\%}$.  The points show the measured
simulation results at each output time, while the lines show the expected
scalings ($\theta^c \propto a^2$ for $n=0$ and $\theta^c \propto a^{4/3}$
for $n=+1$) normalized to the data.}
\label{Mscale.fig}
\end{figure}
Figure~\ref{fM.fig} provides the first evidence that our simulations
are following the expected self-similar scaling laws, at least with
regard to the distributions of masses of collapsed objects.
We now turn to tests that directly examine the scaling of characteristic
group masses, temperatures, Bremsstrahlung luminosities, and densities.

Figure \ref{Mscale.fig} shows the evolution of a fiducial group mass
$\theta^c_{70\%}$, defined so that $70\%$ of the total mass is contained in
groups with $\theta^c \le \theta^c_{70\%}$.  
The ``$c$'' superscript denotes {\it comoving} linear mass densities;
we remove the $a^{-1}$ scaling so that the total ``mass'' $\theta^c_{\rm tot}$
of each species in the simulation box is independent of time.
Isolated particles are counted
as ``groups'' with mass $\Sub{\theta^c}{particle}$ for this purpose.  
We plot the four cooling models ($\hat{t}_C = \infty$, $200$, 1, and 0.1) 
for each initial spectral index $n$,
simulated at our standard
$N = 2 \times 128^2$ particle resolution, and also a high-resolution ($N
= 2 \times 256^2$ particles) simulation of the $n=0$, $\hat{t}_C = 1$ model.
The lines
show the self-similar scaling solutions (a) $\theta^c_{70\%} \propto a^2$ for
$n=0$, and (b) $\theta^c_{70\%} \propto a^{4/3}$ for $n=+1$
(eq. [\ref{Mscale2d.eq}]), normalized to the average amplitude of each
experiment.  This normalization is determined by first scaling each
measurement to a fiducial time (assuming self-similarity), then taking the
average of these scaled measurements.  We plot both the dark matter and
the baryon groups selected with $\ell = 0.2 ~\Delta x_p$ and the
baryon groups only for $\ell = 0.05 ~\Delta x_p$.  There are virtually
no dark matter groups identifiable with such a small linking parameter
because dissipation is required to achieve the necessary overdensity,
at least given our finite mass resolution.

While this Figure (and the succeeding similarly defined figures) are
comparable to the 3-D versions in Paper I, there is one important
computational distinction in how the mass fractions are calculated here.
In the 3-D simulations of Paper I, we typically have between several hundred
and a few thousand identified objects, enough to define a
smooth and well-behaved cumulative mass function.  However, in the 2-D
simulations we typically have an order of magnitude fewer
objects (because we have 16 times fewer particles).
These small numbers make the raw 2-D cumulative mass functions much jumpier
than those in the 3-D simulations, which can make the measured 
$\theta^c_{70\%}$ noisy if we only
use information from those points in the cumulative mass function 
that bracket the 70\% cutoff.  In order to circumvent
this problem, we first fit a polynomial to the cumulative mass
function using general linear least-squares, then interpolate to find the
desired mass fraction cutoff in this smooth polynomial fit.  This 
method allows us
to use information from the entire cumulative mass function in determining
the 70\% cutoff, and we thereby suppress the noise caused by the relatively
small number of collapsed objects.

Clearly all of the models demonstrate good scaling of $\theta^c_{70\%}$
over a range $\log a/a_f \in [-0.8,0]$ for $n=0$ and $\log
a/a_f \in [-1.4,0]$ for $n=+1$, similar to the ranges we expect based on
the resolution limits plotted in Figure \ref{MassRes.fig}.  The scaling
holds equally well for objects identified with $\ell = 0.2 ~\Delta x_p$
(dark matter and baryons) and for dissipated baryon gas objects
selected with $\ell = 0.05 ~\Delta x_p$.  We can see some evidence that the
scalings are poorer at early times, most likely due to the relative lack of
well resolved objects at those times.  
The standard and
high resolution versions of the $n=0$, $\hat{t}_C=1$ model agree
almost perfectly, except that the high resolution simulation is able to
probe back to earlier times.
In both the
$n=0$ and $n=+1$ cases, the baryon objects selected with $\ell = 0.2
~\Delta x_p$ in the models with cooling are slightly more massive than
those in the non-cooling models, suggesting that dissipation plays
some role in concentrating the baryons even at this moderate overdensity.

\begin{figure}[htbp]
\epsscale{0.5}
\plotone{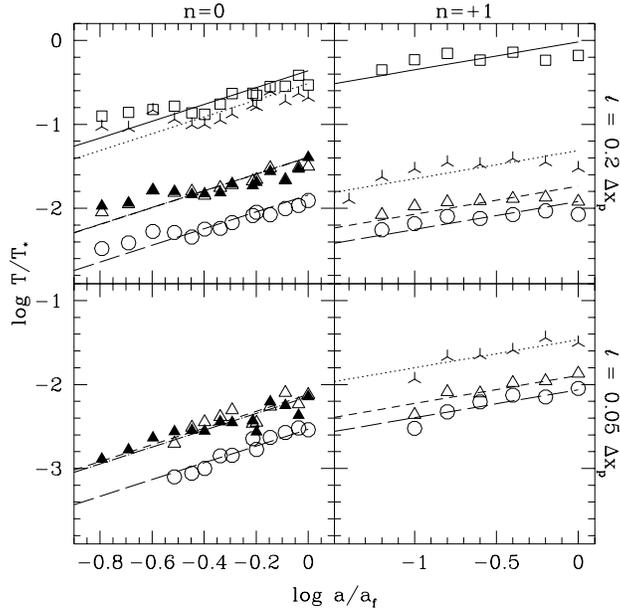}
\epsscale{1}
\caption{Evolution of the emission weighted group temperature $T_{70\%}$
for the $n=0$ (left column) and $n=+1$ (right column) models.  $T_{70\%}$
is defined so that at each expansion factor 
$70\%$ of the mass in the simulation
is contained in groups with $T \le T_{70\%}$.  The lines show the expected
scalings normalized to the data: $T \propto a^1$ for $n=0$ and $T \propto
a^{1/3}$ for $n=+1$.  Point and line types as in Figure
\protect\ref{Mscale.fig}: open squares/solid lines represent the models
without cooling, stellated triangles/dotted lines models with cooling time
$\hat{t}_C=200$, open triangles/short dashes $\hat{t}_C=1$, circles/long
dashes $\hat{t}_C=0.1$, and filled triangles/dot-dash lines the $N = 2
\times 256^2$, $\hat{t}_C=1$ model.}
\label{Tscale.fig}
\end{figure}
In Figure \ref{Tscale.fig} we plot the evolution of a fiducial emission
weighted group temperature $T_{70\%}$, defined in a similar fashion to the
the fiducial mass used in Figure \ref{Mscale.fig}.  The emission weighted
temperature is defined assuming Bremsstrahlung radiation, for which the
bolometric, volume emissivity goes as $\epsilon \propto \rho^2 T^{1/2}$.
This definition
is somewhat inconsistent for the cases with radiative cooling, since
the radiative energy loss from the gas follows our artificial cooling laws
rather than a $T^{1/2}$ law.
We make this choice because the Bremsstrahlung weighted temperature
is closer to an observationally relevant quantity and because
it makes it easier to compare the simulations to each other and to 
the 3-D simulations presented in Paper I.
The emission weighted temperature associated with group $i$ is
\beq
  T_i = \frac{\int \epsilon T ~dA}{\int \epsilon ~dA}
      = \frac{\int \rho^2 T^{3/2} ~dA}{\int \rho^2 T^{1/2} ~dA}
      = \frac{\sum_j \theta_j \rho_j T_j^{3/2}}
             {\sum_j \theta_j \rho_j T_j^{1/2}},
\eeq
represented as a sum over the particles $j$ that are members of group $i$.  We
sort the individual group temperatures in ascending order, and define the
fiducial temperature $T_{70\%}$ such that $70\%$ of the mass is contained
in objects with $T \le T_{70\%}$ (using the same polynomial interpolation
procedure applied to the fiducial mass calculation).
Consistent with the convention adopted
for the mass scaling test shown in Figure \ref{Mscale.fig}, we count the
mass contained in unresolved groups (or individual particles) as having
temperatures less than the lowest measured group temperature.  The lines
show the self-similar solution normalized to each experiment.  Temperatures
are scaled to the hydrostatic equilibrium support temperature 
(eq.~[\ref{Tvc2d.eq}]) of a $\theta_*$ object at the final expansion factor
for each experiment: $T_*(a_f) = G \mu m_p (2 k_B)^{-1} \theta_*(a_f)$, 
where $\theta_*(a_f)$ is the mass scale associated with the density 
contrast $\delta_*$ that fits the $\ell = 0.2 ~\Delta x_p$
mass function, as described in \S \ref{Sims.sec}.  
Note that if we were to scale instead to the characteristic temperature at 
each expansion factor $T_*(a)$,
then the predicted evolution tracks would be horizontal lines.

In Figure \ref{Tscale.fig} we see that the fiducial emission weighted
temperature scales well in nearly all of the simulations, both for
$\ell = 0.2 ~\Delta x_p$ and for $\ell = 0.05 ~\Delta x_p$.  
The logarithmic slope of the scaling relation is three times higher
for the $n=0$ models than for $n=+1$ ($T \propto a^1$ vs. $T \propto a^{1/3}$),
and the simulations clearly capture this distinction (note, however,
that we adapt 
the range of the horizontal axis to the dynamic range of the experiments).
The standard and high resolution simulations of the $n=0$,
$\hat{t}_C=1$ model yield nearly identical results.

Figure~\ref{Tscale.fig} clearly shows
the effects of radiative cooling on the typical gas temperatures: 
the collapsed gas in the $\hat{t}_C=0.1$, $n=0$ model, 
for example, is 1.5 dex cooler than in the case without
cooling.  Additionally, for the models with cooling, the dense, dissipated
gas selected with $\ell = 0.05 ~\Delta x_p$ is typically 0.5 dex cooler
than the $\ell = 0.2 ~\Delta x_p$ gas.  As noted in the mass scaling tests,
we again see poorer scaling at early times, particularly noticeable in the
$n=0$, $\ell = 0.2 ~\Delta x_p$ objects, where we only find reasonable
scaling for $\log a/a_f \in [-0.6,0]$.  The dissipated, 
$\ell = 0.05 ~\Delta x_p$ objects do not even form until the larger
$\ell = 0.2 ~\Delta x_p$ structures begin to scale well, but once this happens
the temperatures of the dissipated groups always scale well.  For the $n=+1$
models, the temperatures of the dissipated structures in the lower panel
always scale more effectively than those of the larger $\ell = 0.2 ~\Delta
x_p$ groupings.  This difference suggests that the cooled gas cores of the
structures in the $n=+1$ models scale correctly, and that it is the hot,
diffuse gas in the outer regions that does not.  This result seems contrary
to what we find in the 3-D models without cooling in Paper I, and it could be
due to an increased importance of hot gas that is not physically bound
being erroneously included in collapsed structures by
friends-of-friends with the larger linking parameter.  We will 
consider this question further in \S \ref{CoolGas.sec}.

\begin{figure}[htbp]
\plottwo{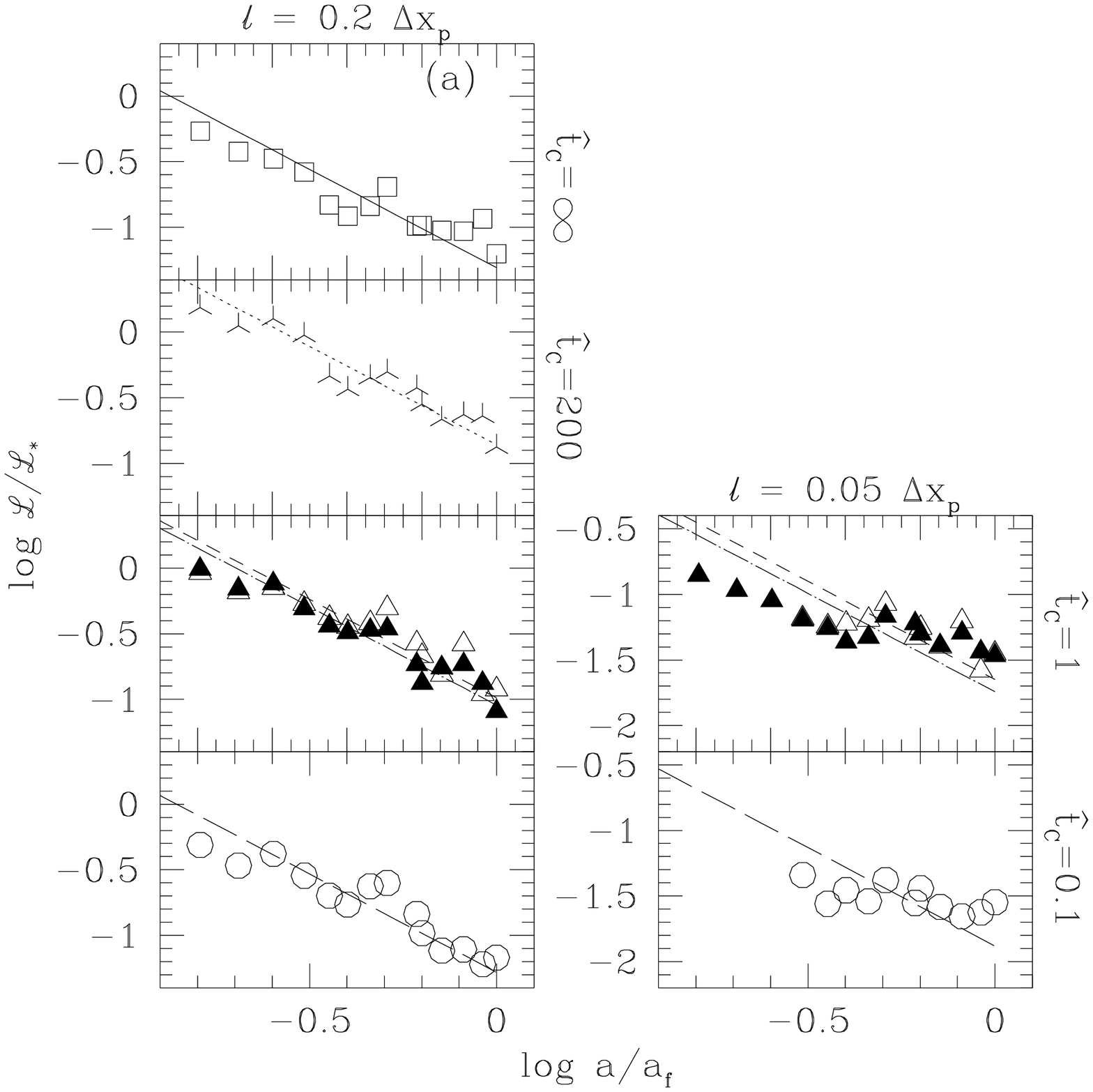}{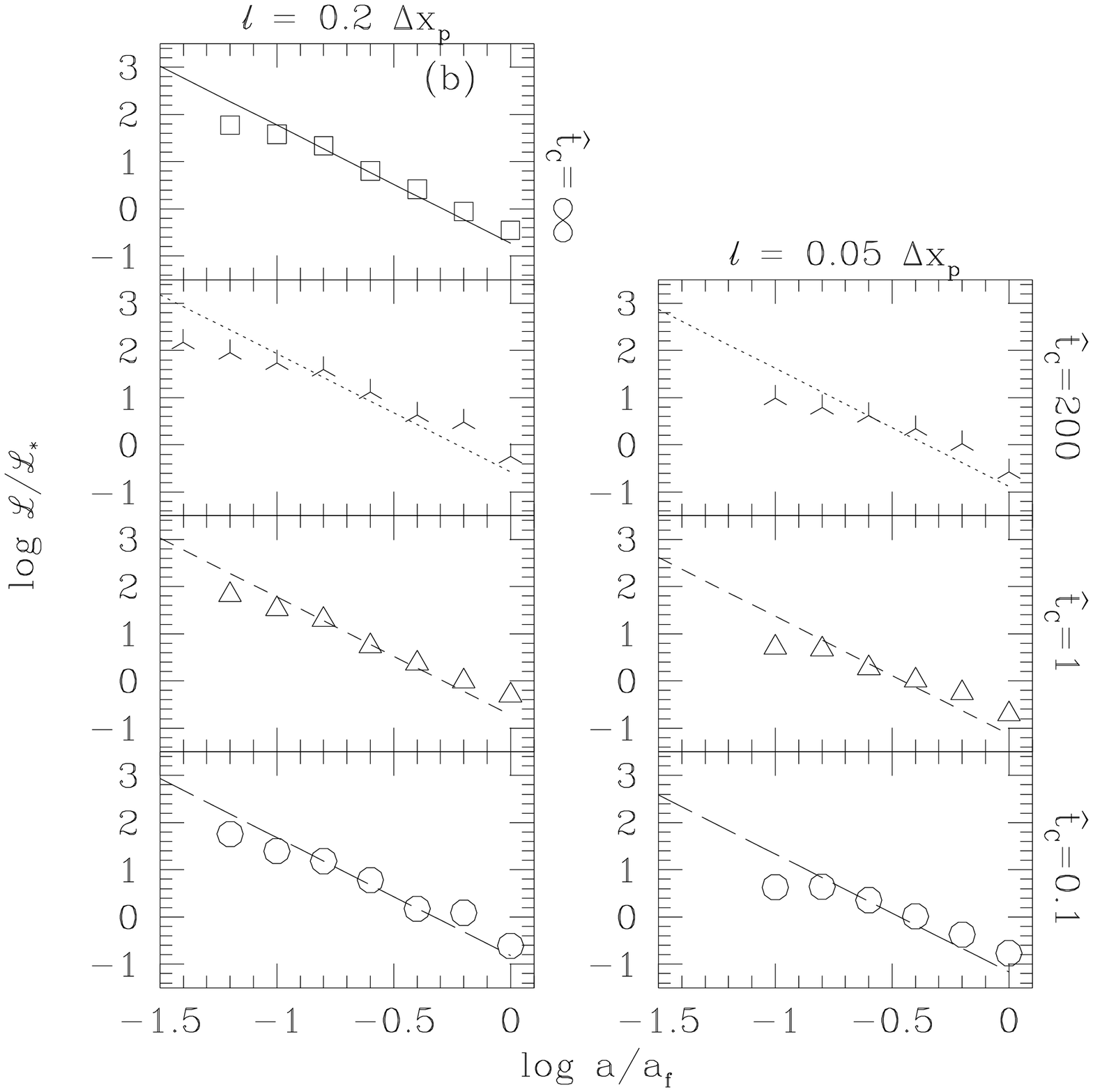}
\caption{Evolution of the group luminosity $\cL_{70\%}$ for the (a) $n=0$
and (b) $n=+1$ models, such that at each expansion $70\%$ of the mass in the
simulation is contained in groups with $\cL \le \cL_{70\%}$.  Note that
the ``luminosity'', $\cL$, is in fact a luminosity per unit length.  The
lines show the expected scalings normalized to the data: $\cL \propto
a^{-3/2}$ for $n=0$ and $\cL \propto a^{-5/2}$ for $n=+1$.}
\label{Lscale.fig}
\end{figure}
Figure \ref{Lscale.fig} tests the scaling of a fiducial group luminosity
$\cL_{70\%}$, defined in much the same manner as the fiducial group
temperature in Figure \ref{Tscale.fig}.  For each group the total
luminosity is defined as $\cL = \int \rho^2 T^{1/2} ~dV = \sum_j \theta_j
\rho_j T_j^{1/2}$.  As with the temperature, we sort the groups in order of
increasing luminosity and find the luminosity $\cL_{70\%}$ such that
$70\%$ of the total baryon mass is contained in groups with $\cL \le
\cL_{70\%}$.  The luminosities are scaled to the characteristic luminosity
at the final expansion factor for each model, $\cL_*(a_f) = \theta_*(a_f)
~\rho_*(a_f) ~T^{1/2}_*(a_f)$, where we again use quantities appropriate
for the $\ell = 0.2 ~\Delta x_p$ groups.  The lines show the self-similar
predictions $\cL \propto a^{-3/2}$ for $n=0$ and $\cL \propto a^{-5/2}$
for $n=+1$ (eq. [\ref{Lscale2d.eq}]).  The luminosity does not depend
strongly on the adopted cooling law because the reduction in temperature
in models with stronger cooling is compensated by the increase in gas density.

Relative to the poor scaling of group luminosities found for 3-D objects
in Paper I, the scaling behavior in Figure~\ref{Lscale.fig} is remarkably good.
The $\ell = 0.2 \Delta x_p$ groups follow the analytic scaling laws for
all combinations of $n$ and $\hat{t}_C$.  The dissipated, $\ell=0.05$
groups show some tendency to have lower luminosities at earlier times
(when the groups are less well resolved), but even in these cases the
luminosity scaling is not bad.  Furthermore, the high and low resolution
versions of the $n=0$, $\hat{t}_C=1$ model concur almost perfectly, in
marked contrast to the strong resolution dependence found for group
luminosities in the 3-D non-cooling models of Paper I.  We will return to
consider the difference between our present and previous results, following 
a discussion of the gas density scaling tests.

\begin{figure}[htbp]
\plottwo{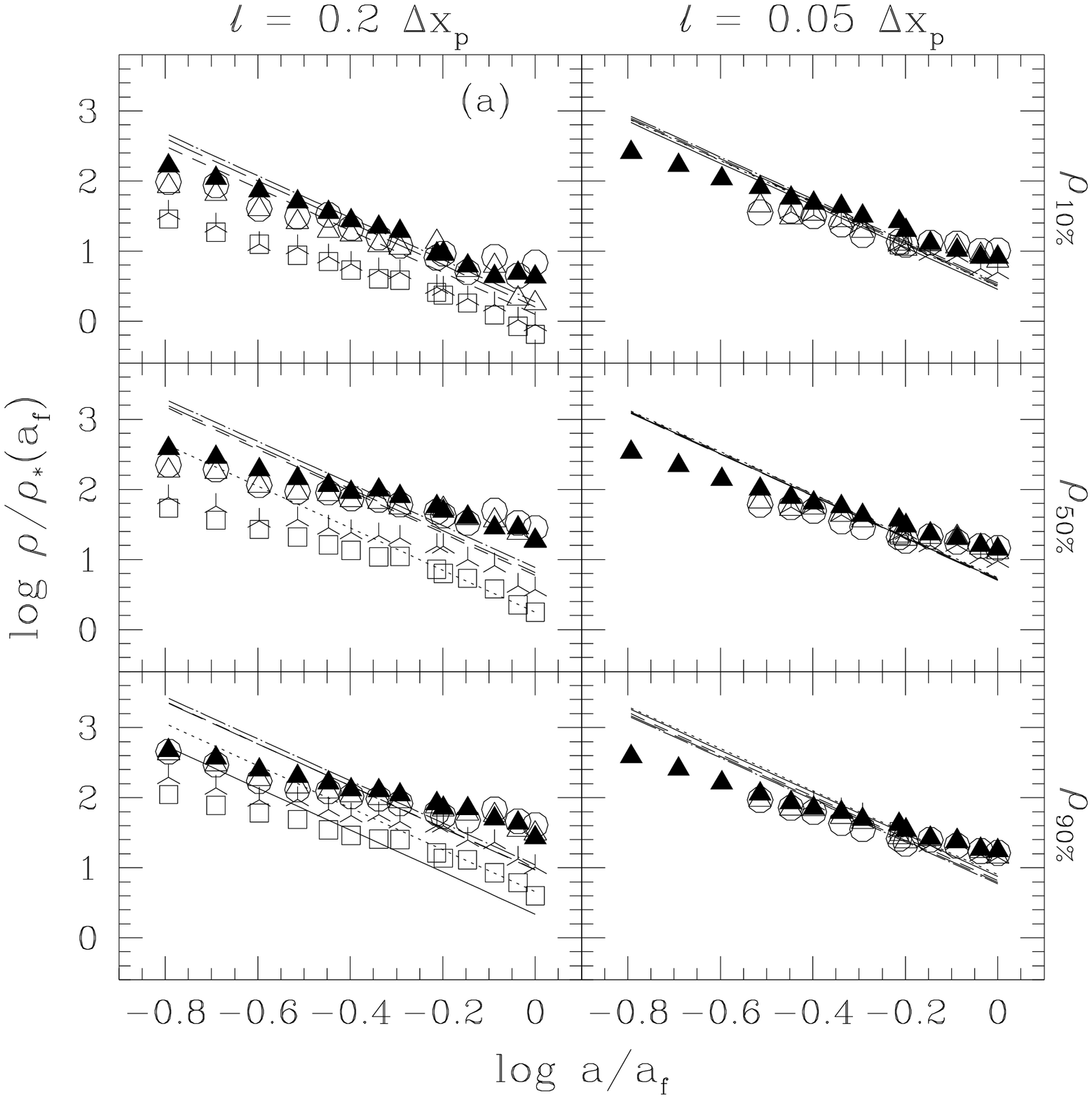}{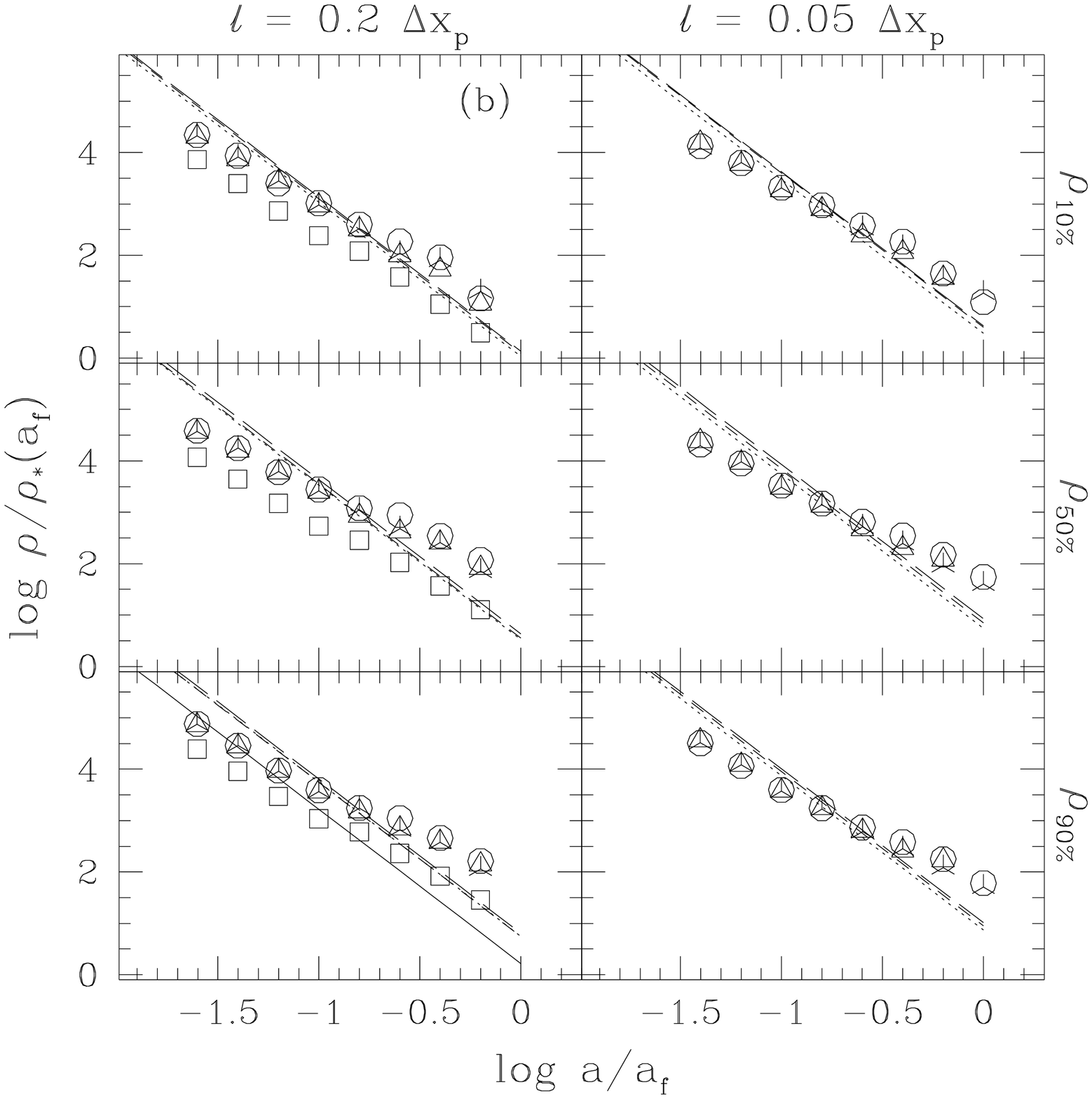}
\caption{Evolution of the group density $\rho_{x\%}$ for the (a) $n=0$
and (b) $n=+1$ models.  $\rho_{x\%}$ is defined as the average of
$\rho_{x\%}^i$ for groups in a given range of $\theta$, where
$\rho_{x\%}^i$ for each group $i$ is the density such that $x\%$ of the
mass in the group is at densities $\rho \le \rho_{x\%}^i$.  In each figure,
the left column shows the results for groups selected with linking
parameter $\ell = 0.2 ~\Delta x_p$ and the right column for groups selected
with $\ell = 0.05 ~\Delta x_p$.  The rows show the averages for
$\rho_{10\%}$, $\rho_{50\%}$, and $\rho_{90\%}$, progressively sampling
from the outskirts of each group inward toward the core.  For the $n=0$
models in part (a) we average over objects in the mass range $0.5
~\protect\Sub{\theta}{nl} \le \theta \le 2.5 ~\protect\Sub{\theta}{nl}$,
while for the $n=+1$ models in part (b) we average over groups in the range
$0.2 ~\protect\Sub{\theta}{nl} \le \theta \le 1.0
~\protect\Sub{\theta}{nl}$.  The lines show the expected scaling
normalized to the data: $\rho \propto a^{-3}$.  Point and line types as
defined in Figure \protect\ref{Mscale.fig}: open squares/solid lines
represent the models without cooling, stellated triangles/dotted lines
models with cooling time $\hat{t}_C=200$, open triangles/short dashes
$\hat{t}_C=1$, circles/long dashes $\hat{t}_C=0.1$, and filled
triangles/dot-dash lines the $N=2 \times 256^2$, $\hat{t}_C=1$ model.}
\label{rhoscale.fig}
\end{figure}

We cannot test the gas density scaling in
precisely the same manner as the previous quantities because groups are
effectively selected by their overdensities, and we might therefore expect
their average densities to scale by construction.  We approach this problem 
as we did in Paper I, by
first calculating a percentile density $\rho_{x\%}$ for each group, such
that $x\%$ of the gas particles in the group are at densities $\rho_j \le
\rho_{x\%}$.  We then take all of the groups in a given mass range and find
the average value of $\rho_{x\%}$, where the average is weighted by the
group mass, with the constraint that a group must contain at least 28
particles to contribute.  In Figure \ref{rhoscale.fig} we plot average
densities defined in this way for $\rho_{10\%}$, $\rho_{50\%}$, and
$\rho_{90\%}$, progressively probing from the outskirts of each group
inward.  The left columns are calculated for objects selected with the
larger linking parameter $\ell = 0.2 ~\Delta x_p$ (including both the hot
and cold components of collapsed structures), while the right columns use
the smaller linking parameter $\ell = 0.05 ~\Delta x_p$ (effectively 
restricting the
selection to only the cooled gas cores in the simulations with radiative
cooling).

Considering first the $\ell = 0.2 ~\Delta x_p$ groups in the left columns,
the effects of cooling are quite evident, as models with stronger cooling
show markedly higher densities.  We can also see that the behavior of the
density noted in Paper I is qualitatively reproduced here (most notably in
the $n=0$ models): the lower density cuts (probing the outer
regions of the objects) scale most effectively, but as we consider
progressively higher density cuts (and therefore move inward to smaller
radii) the gas density progressively scales more poorly.  
For $n=+1$ the density scaling is better, and the differences in the scaling
behavior of $\rho_{10\%}$ and $\rho_{90\%}$ are smaller, though they
follow the same trend.
The density evolution for the dissipated baryon structures selected with
$\ell = 0.05 ~\Delta x_p$ (the right columns in Figure \ref{rhoscale.fig})
is similar to $\rho_{90\%}$ in the $\ell = 0.2 ~\Delta x_p$ objects
and demonstrates relatively little change between $\rho_{10\%}$,
$\rho_{50\%}$, and $\rho_{90\%}$.  Presumably this similarity arises because the
dissipated groupings selected with $\ell = 0.05 ~\Delta x_p$ are formed at
the cores of the larger, less dense structures identified by $\ell = 0.2
~\Delta x_p$, and therefore $\rho_{90\%}$ in $\ell = 0.2 ~\Delta x_p$
probes the same material selected with the smaller linking parameter.

Comparing these 2-D results to the 3-D results of Paper I, we find that
the core densities ($\rho_{90\%}$) scale more accurately in 2-D but
that the halo densities ($\rho_{10\%}$) scale somewhat less accurately.
This difference most likely reflects the interplay between the dimensionality 
of the models and the Lagrangian resolution of SPH, in which quantities
are averaged over fixed numbers of neighbors.  In both 2-D and 3-D 
simulations we find that the average radial density profiles approximately
follow $\rho(r) \propto r^{-2}$ over the scales we resolve.  
In 2-D, the number of particles in a
radial shell goes as $\Sub{N}{2-D}(r) \propto \rho(r) ~dA = 2 \pi r \rho(r)
~dr \propto r^{-1}$, while in 3-D this becomes $N(r) \propto \rho(r) ~dV =
4 \pi r^2 \rho(r) ~dr \propto r^{0}$.  
Thus, for 2-D objects we tend to have relatively poorer
resolution for large $r$ (the region probed by $\rho_{10\%}$), and better
resolution for small $r$ ($\rho_{90\%}$).  

The better scaling of core gas densities in these 2-D experiments 
partly accounts for the improved scaling of the luminosities 
relative to their 3-D counterparts in Paper I.
However, the core gas densities do not scale perfectly, and this
change alone seems unlikely to account for so much improvement in the
luminosity scaling.  The change from traditional SPH to ASPH also helps ---
we have repeated some of the 2-D tests with SPH and find that the 
luminosity scaling is somewhat worse than with ASPH but is still markedly
better than that found in Paper I.  The primary cause of the improved
scaling, however, is that the luminosities of the 2-D groups are less
core dominated and are therefore less sensitive to poorly resolved gas.
The 2-D groups have nearly isothermal temperature profiles, while the
3-D temperature profiles fall with radius
(roughly following $T \propto r^{-0.6}$).  
Since the 2-D and 3-D density profiles are nearly the same
($\rho \propto r^{-2}$), the flatter temperature profile in 2-D 
leads to a less sharply peaked luminosity profile.

\section{Scaling in the Cooled Baryon Component}
\label{CoolGas.sec}
Of all the questions we can ask of our simulations, the one that goes
to the heart of the galaxy formation issue is this: do they correctly
follow the collapse and dissipative settling of baryons into dense,
radiatively cooled objects?
The scaling of the mass function (Figure~\ref{fM.fig}) and other
properties of high density ($\ell=0.05 \Delta x_p$) groups already
suggests that the answer to this question is at least a qualified yes.
In this Section, we focus more directly on the cooled baryon component
of virialized groups.

\begin{figure}[htbp]
\epsscale{0.5}
\plotone{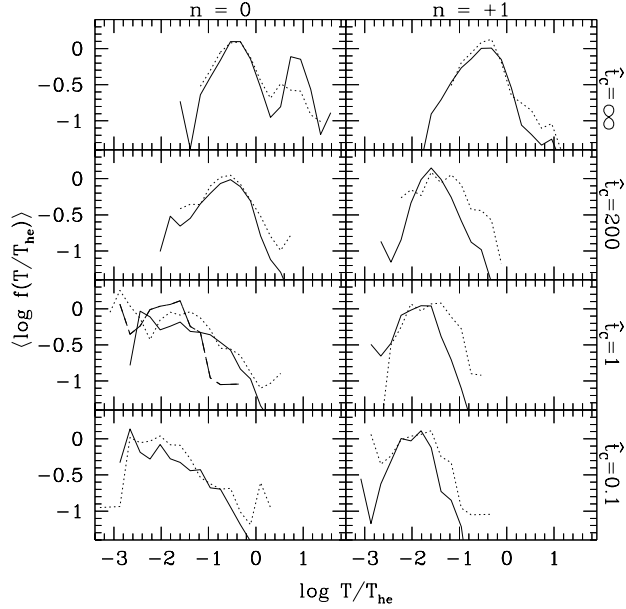}
\epsscale{1}
\caption{Average gas temperature distributions for groups in
the $n=0$ (left column) and $n=+1$ (right column) simulations.
Groups are identified by applying the friends-of-friends algorithm
with $\ell = 0.2 ~\Delta x_p$ to the full particle
distributions (baryons and dark matter).
Temperatures are scaled to $\Th = \mu m_p (2 k_B)^{-1} G \theta$,
where $\theta$ is the total linear mass density of the group.  The solid
curves are measured for groups at $a/a_f = 1$ and the dotted curves for
groups at $a/a_f = 0.4$.  In the $n=0$, $\hat{t}_C=1$ panel we also plot the
high resolution, $N=2 \times 256^2$ simulation at $a/a_f = 1$ with the short
dashed lines and at $a/a_f = 0.4$ with long dashes.  Since all quantities in
this plot are defined in a scale-free manner, the profiles measured at
different times should lie on top of one another if these structures are
correctly obeying the expected self-similar behavior.  In order to
calculate these average distributions, the baryon mass in each group is
binned as a function of $T/\Th$, and the individual distributions for
groups within a given range of $\theta$ ($0.5 ~\protect\Sub{\theta}{nl}
\le \theta \le 2.5 ~\protect\Sub{\theta}{nl}$ for $n=0$, and $0.2
~\protect\Sub{\theta}{nl} \le \theta \le 1.0 ~\protect\Sub{\theta}{nl}$ for
$n=+1$) are then averaged together.}
\label{TtoTcprof.fig}
\end{figure}

We begin by running
friends-of-friends with linking parameter $\ell = 0.2 ~\Delta x_p$ on {\em
all} particles in a given experiment, yielding structures comprised of both
baryons and dark matter (while up to now we have considered baryon
and dark matter groups separately).  Equation~(\ref{Tvc2d.eq}) 
gives a characteristic support temperature for an
object based purely on its mass and the assumption of hydrostatic
equilibrium: $\Th = \mu m_p (2 k_B)^{-1} G \theta$.  In Figure
\ref{TtoTcprof.fig}, we plot average baryon temperature distribution
functions for objects that fall within a characteristic mass range, where
the temperatures are scaled to this characteristic temperature for each
structure.  The average profiles are determined by first computing
individual distributions for the baryon mass $f_i(T/\Th)$ for each group
$i$, then averaging these individual profiles in a restricted range of
$\theta$, chosen so that we will have a number of resolved objects in that
mass range at the times we consider.

Each panel of Figure \ref{TtoTcprof.fig} shows average
temperature distributions calculated at two different expansion
factors ($a/a_f = 0.4$ and 1). Since the temperatures are
scaled by the group hydrostatic equilibrium temperature
and the mass range of groups contributing to the average distribution
is scaled to $\Sub{\theta}{nl}$, the two curves should lie on top
of one another if the simulations perfectly obey the
expected self-similar scalings.
The agreement between the dotted and solid curves in Figure
\ref{TtoTcprof.fig} shows that the temperature distribution of
material, both dissipated and shocked, is accurately following self-similar
evolution.  The high resolution and standard resolution simulations
of the $n=0$, $\hat{t}_C=1$ model also correspond well, suggesting that this
distribution is insensitive to numerical resolution effects.  As one would
expect, the models with stronger cooling show an increased fraction of the
baryons at cooler temperatures.  As we increase the strength of
the cooling in the $n=0$ models the distribution of gas temperatures
becomes wider, developing a more extended low temperature tail that
contains the bulk of the gas.  In the $n=+1$
models the temperature distribution maintains
a bell shape and simply shifts toward lower
mean temperature with increased cooling.  
For $\hat{t}_C \leq 200$, the temperature distributions are also
less sensitive to the cooling amplitude
in the $n=+1$ models than in the $n=0$ models.

For both $n=0$ and $n=+1$, we find that some gas substantially
hotter than $\Th$ is incorporated into the $\ell = 0.2 ~\Delta x_p$
groups, especially in the models without cooling.  While this hot gas
must be shocked, it may represent a different population (presumably
hot, diffuse gas on the outskirts of these structures) from the
interior, cooler gas at $T \lesssim \Th$.  This extra population
of hot gas may help explain the surprising
behavior noted in our discussion of Figure \ref{Tscale.fig},
where we find that the emission weighted temperatures of objects identified
with the smaller linking parameter actually scale better than those found
with the large $\ell$, since the
objects selected with the smaller linking parameter do not include this
extra gas.  

In all cases the temperature
distributions in Figure~\ref{TtoTcprof.fig}
are fairly continuous, while in typical, realistic cold dark
matter (CDM) simulations (see, \eg, Katz, Hernquist, \& Weinberg 1992 or
Evrard, Summers, \& Davis 1994) the temperature distribution of gas in such
objects is more nearly bimodal, with
dissipated gas having cooled efficiently to $T \sim 10^4$K from shocked
temperatures of $T \sim 10^5-10^6$K.  Though we do not show the 
results here, we have
repeated our 2-D models using a realistic cooling law such as that used in Katz
\etal\ (1992), and in such cases we recover the typical bimodal temperature
distribution, with cooled, dense gas knots embedded in hot, low density gas
halos.  This test shows that it is our featureless power-law cooling
relations that cause the temperature distributions to have the
continuous form seen in Figure~\ref{TtoTcprof.fig}, not the difference
in dimensionality.

We are especially interested in the fraction of the baryons that
cool to low temperatures following collapse and shock heating in
dark matter halos.  In full-scale simulations of galaxy formation,
it is these radiatively cooled baryons that are converted into stars,
creating the luminous regions of observable galaxies.
If cosmological simulations are to correctly predict galaxy
luminosity functions, star formation rates, and so forth, they must
first correctly calculate the masses of the cold baryon concentrations
that form the galaxies.

\begin{figure}[htbp]
\epsscale{0.5}
\plotone{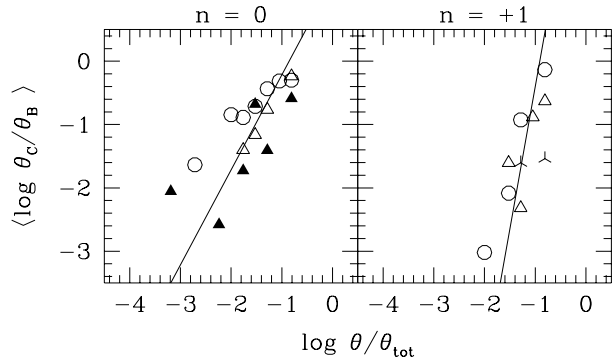}
\epsscale{1}
\caption{Average fraction of baryon mass per group with $T/\Th \le 0.005$
as a function of total group $\theta$ at $a=a_f$ for the $n=0$ (left) and
$n=+1$ (right) simulations.  The lines show the expected relations
$\theta_C/\theta_B \propto \theta^{3/2}$ ($n=0$) and $\theta_C/\theta_B
\propto \theta^{9/2}$ ($n=+1$), based on the assumptions that objects of
differing masses are similar and that the cooled mass is inversely
proportional to
cooling time, as described in the text.  Point types as defined in Figure
\protect\ref{Mscale.fig}: stellated triangles for models with cooling time
$\hat{t}_C=200$, open triangles $\hat{t}_C=1$, circles $\hat{t}_C=0.1$, and
filled triangles the $N=2 \times 256^2$, $\hat{t}_C=1$ model.}
\label{TtoTcfM.fig}
\end{figure}

We will turn to the self-similar scaling of the cooled baryon
fractions shortly, but we first examine the dependence of this
fraction on total mass at fixed time.  Figure~\ref{TtoTcfM.fig}
plots $\langle \theta_C/\theta_B \rangle$ as a function of
total group mass $\theta$ at the final output time, where
$\theta_C/\theta_B$ is the fraction of baryons in the group that
have cooled to $T \leq 0.005\Th$, and the averages are computed
in logarithmic bins of $\theta$.  We choose a low temperature
threshold so that we probe only regions that have undergone radiative cooling,
and as a result only the strong cooling models
($\hat{t}_C=1$ and $\hat{t}_C=0.1$) show up on the plot,
along with two points for $\hat{t}_C=200$, $n=+1$.
The $\hat{t}_C=0.1$ points (circles) show a clear trend of
increasing $\theta_C/\theta_B$ with increasing $\theta$,
with a steeper correlation for $n=+1$ than for $n=0$.
The $\hat{t}_C=1$ points (triangles) have larger scatter
but appear to show a similar trend, especially with the higher
dynamic range afforded by the $2\times 256^2$, $n=0$ simulation.

These trends are expected given the temperature dependence
of our cooling functions.  The friends-of-friends algorithm
selects groups with similar overdensity at all $\theta$, and
equation~(\ref{Tvc2d.eq}) predicts that the characteristic
temperature of a group in the absence of cooling is proportional to $\theta$.
The cooling time is proportional to $\rho T /\Lambda(T) \propto T^{1-\beta}$
for groups of similar $\rho$, where $\beta$ is the index of the
cooling function given by equation~(\ref{beta2d.eq}).
For $\beta>1$ this argument implies a shorter cooling time in more
massive groups.  The lines in Figure~\ref{TtoTcfM.fig} show
the relations
\beq
  \label{CoolMass.eq}
  \frac{\theta_C}{\theta_B} \propto \hat{t}_C^{-1} 
                            \propto T^{\beta - 1} 
                            \propto \theta^{\beta - 1},
\eeq
i.e., a cooled baryon fraction inversely proportional to the
cooling time.  While the relation between cooled fraction and
cooling time need not be this simple, these lines do seem to
capture the numerical results fairly well, explaining both
the steep trend of cooling with mass and the marked difference
in slope between $n=0$ and $n=+1$.

Since a given group has gas at a range of densities and temperatures,
even in the absence of cooling, our argument above makes the implicit
assumption that groups of different mass have similar density and
temperature profiles in scaled units before the onset of cooling:
formally speaking, $\rho/\bar{\rho}$ and $T/\Th$ have the same
functional dependence on $r/r_{\rm vir}$, where $r_{\rm vir}$ is
the group virial radius.  This type of similarity is {\it not}
guaranteed by our choice of scale-free initial conditions, which only 
ensures self-similar time evolution.  (For example, scale-free
physics implies that a $\theta_*$ group at time $t_1$ is similar
to a $\theta_*$ group at time $t_2$, but it does not imply that
a $0.5\theta_*$ group is similar to a $\theta_*$ group at fixed time.)
This assumption is justifiable in light of recent
N-body experiments, such as those of Navarro, Frenk \& White (1995, 1996,
1997) or Cole \& Lacey (1996), who find that collapsed structures in
high-resolution N-body experiments tend to a universal density profile over
roughly two orders of magnitude in mass.  We have directly examined the
objects in our experiments and verified that the density and temperature
profiles are similar over the mass ranges we resolve.

\begin{figure}[htbp]
\plottwo{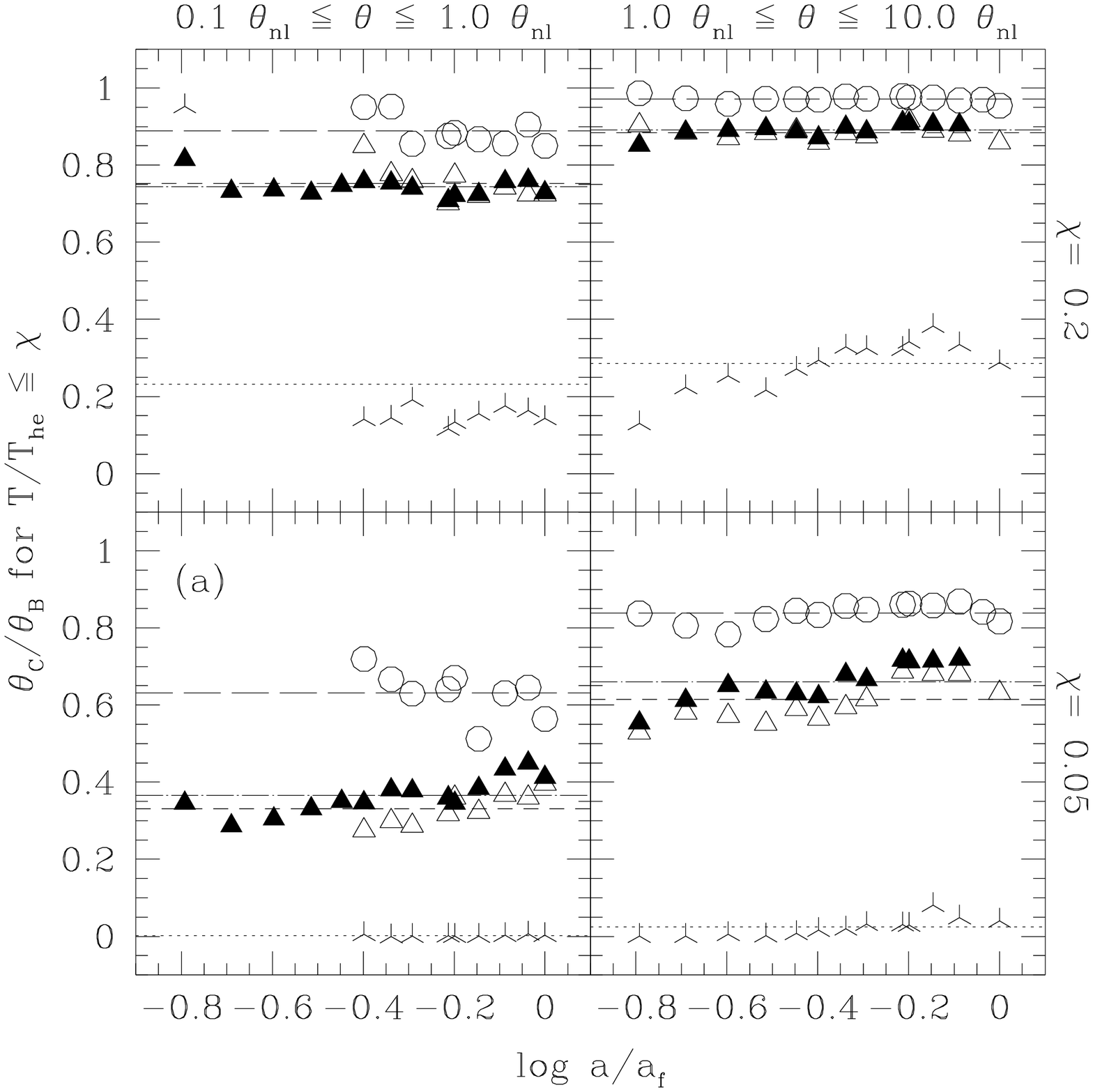}{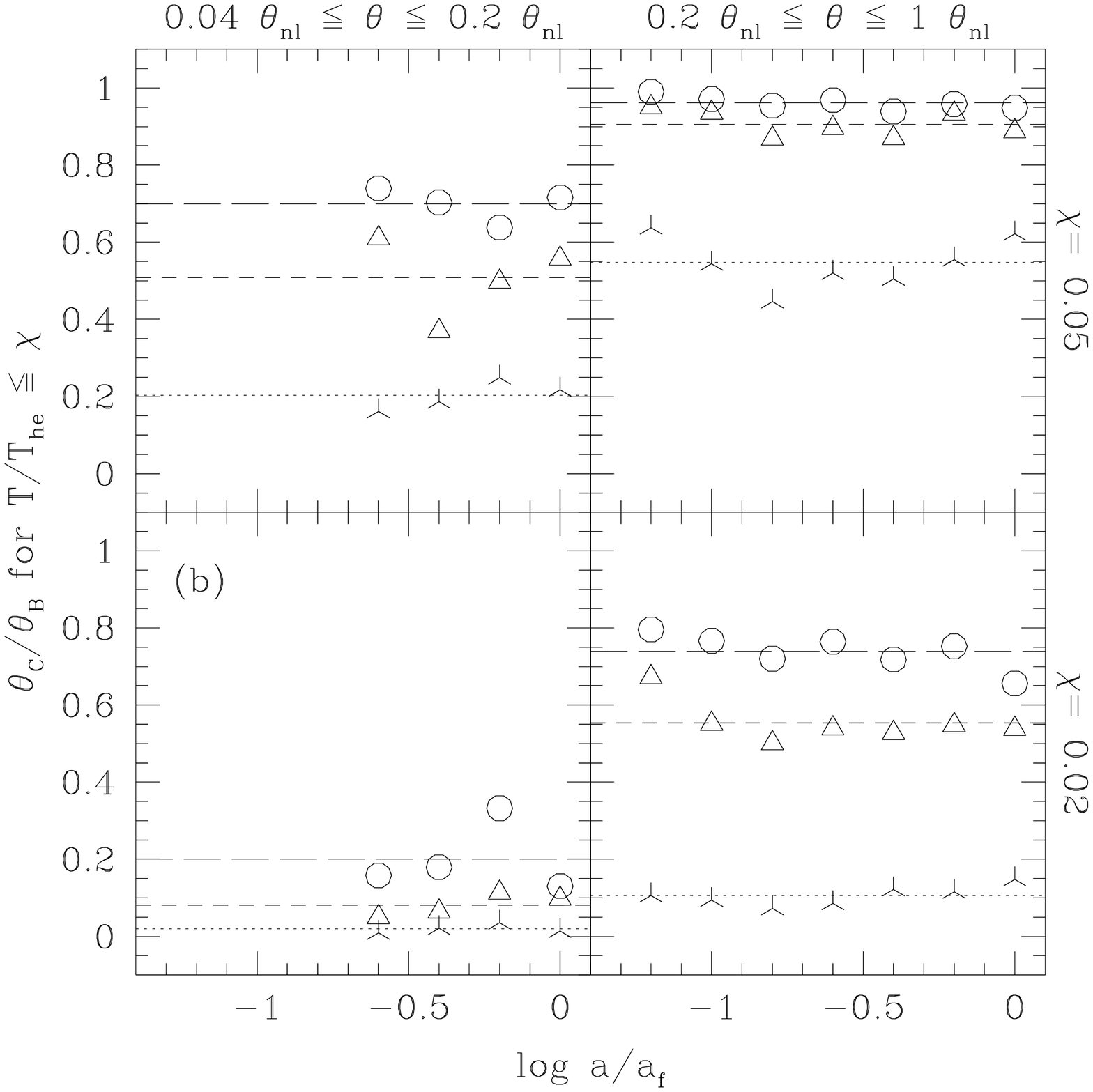}
\caption{Average fraction of baryon mass in groups with $T/\Th \le \chi$ as
a function of expansion factor
for the (a) $n=0$ and (b) $n=+1$ models.  At each expansion factor,
the fraction of baryon mass $\theta_C/\theta_B$
with $T/\Th \le \chi$ is calculated for each group
(where $\chi$ is given in the right
side row labels), then averaged for all groups in a given range
of $\theta$ (given in
column headings).  Since all quantities in this figure are defined in a
scale-free manner, the cooled baryon mass fraction is expected to remain
fixed throughout the evolution, as shown by the lines.  Point and line
types as defined in Figure \protect\ref{Mscale.fig}: stellated
triangles/dotted lines for models with cooling time $\hat{t}_C=200$, open
triangles/short dashes $\hat{t}_C=1$, circles/long dashes $\hat{t}_C=0.1$,
and filled triangles/dot-dash lines the $N=2 \times 256^2$, $\hat{t}_C=1$
model.}
\label{Mcoolfrac.fig}
\end{figure}
Figure \ref{Mcoolfrac.fig} directly addresses the central issue of
our investigation, the scaling of the cooled baryon mass fraction.
We take each object identified at
a given output time and calculate the fraction of the baryon mass 
that has cooled below a fixed fraction of the hydrostatic equilibrium
temperature, $T/\Th \le \chi$.  We then compute the average of
this mass fraction for all groups in a given range of group
mass and plot these values as a function of the expansion factor.  

The upper left panel of Figure \ref{Mcoolfrac.fig}a 
shows the average fraction of gas with $T/\Th \le 0.2$, for groups
with mass in the range $0.1\Sub{\theta}{nl}$ to $\Sub{\theta}{nl}$.
As expected, this fraction is higher for models with shorter cooling
times: 15\%, 75\%, and 90\% for $\hat{t}_C=200$, 1, and 0.1, respectively.
While self-similarity does not tell us what this fraction should
be for a given model, it does tell us that the fraction should be
independent of time, since the ratio $T/\Th$ is dimensionless and the
mass range is scaled in terms of $\Sub{\theta}{nl}$.
The simulation results agree with this analytic scaling law remarkably
well, with cooled gas fractions that are independent of expansion
factor in the range $\log a/a_f \in [-0.4,0]$.
The standard and high resolution simulations of the $\hat{t}_C=1$ model are
in in nearly perfect agreement for $\log a/a_f > -0.4$, and the
high resolution simulation extends the scaling back to $\log a/a_f = -0.8$.
If numerical resolution effects were important, we would expect to
find a difference between these two simulations, and we would expect the
cooled gas fraction to increase with time in any given simulation
as groups in the range $0.1\Sub{\theta}{nl} - \Sub{\theta}{nl}$
are resolved by progressively more and more particles.

The lower left panel of Figure \ref{Mcoolfrac.fig}a shows the average fraction
of mass in these groups with $T/\Th \leq 0.05$.  While less gas has cooled
below this lower temperature threshold, the cooled gas fractions still show
the analytically predicted scaling.  The more massive groups have more
cooled gas (right column of Figure~\ref{Mcoolfrac.fig}a), as expected
on the basis of Figure~\ref{TtoTcfM.fig} and the accompanying discussion.
The cooled gas fractions scale well over a larger range of expansion
factor, $\log a/a_f \in [-0.8,0]$, because these more massive groups
are well resolved at earlier times.

Figure \ref{Mcoolfrac.fig}b shows the corresponding results for the
$n=+1$ models.  We select different mass ranges that correspond better
to the numerical limits of the $n=+1$ simulations 
(see Figure~\ref{MassRes.fig}), and we adopt lower temperature 
thresholds because cooling is stronger in the $n=+1$ models
(see Figure~\ref{TtoTcprof.fig}).
The cooled baryon fractions of the $n=+1$ groups scale well for 
$\log a/a_f \in [-0.6,0]$ in the mass range 
$0.04\Sub{\theta}{nl}-0.2\Sub{\theta}{nl}$ and for
$\log a/a_f \in [-1.0,0]$ for
$0.2\Sub{\theta}{nl}-1.0\Sub{\theta}{nl}$.

In the simulations of Paper I, with no cooling, we found poor
scaling of group Bremsstrahlung luminosities.  We attributed
this failure to poor scaling of the group's core densities ---
as groups are resolved by progressively more particles, the
estimated central densities increase because of better resolution.
Sensitivity of the gas density to numerical resolution has also been noted in a
number of previous investigations, including Kang \etal\ (1994), Anninos \&
Norman (1996), Frenk \etal\ (1996), Weinberg, Hernquist, \& Katz (1997),
and Owen \& Villumsen (1997), to name merely a few.  

Since radiative cooling is itself sensitive to gas density
($\Lambda \propto \rho^2$), we initially expected that it would
be difficult to find self-similar scaling in simulations with
radiative cooling.  How, then, do we account for the nearly perfect scaling 
of the cooled baryon fractions in Figure~\ref{Mcoolfrac.fig}?
The key to the explanation is that the cooling instability is
effectively a threshold phenomenon.  If the numerical resolution in an
object is insufficient, so that the cooling time at the highest 
resolved density is longer than the dynamical time, then radiative 
cooling is almost entirely suppressed.  Once the resolved density passes
a critical threshold, however, thermal instability sets in, and the
gas rapidly cools, loses pressure support, and converges toward
the physical solution.
The fact that these simulations do not form significant amounts
of radiatively dissipated gas until the larger objects begin to
demonstrate the expected self-similar scalings
(as noted in our discussion of the mass and temperature scaling
tests in Figures~\ref{Mscale.fig} and~\ref{Tscale.fig}),
indicating that the physical solution is beginning to be followed
correctly, supports this interpretation.
The agreement between the standard and high resolution simulations
of the $n=0$, $\hat{t}_C=1$ model and the successful scaling of the
mass function of dissipated objects (Figure~\ref{fM.fig}),
the temperature distributions in collapsed groups (Figure~\ref{TtoTcfM.fig}),
and the cooled baryon fractions (Figure \ref{Mcoolfrac.fig})
all indicate that the simulations have converged to a physical
solution that is, at least for these measures, insensitive to their
numerical resolution.

\section{Summary and Conclusions}
We analyze a set of 2-D hydrodynamic simulations designed to study the
self-similar evolution of hierarchical structure in the presence
of radiative cooling.  The radiative cooling law for a 
primordial H/He plasma would introduce a preferred timescale in these
models, so we instead use artificial cooling laws 
that maintain the scale-free nature of the physics: 
$\Lambda(T) \propto T^\beta$ where $\beta$ is a function
(eq.~[\ref{beta2d.eq}]) of the spectral index $n$ of the 
initial density fluctuations.
We simulate eight distinct physical models, with $n=0$ and $n=+1$
initial power spectra and four different
amplitudes of the corresponding radiative
cooling laws, ranging from no cooling to very strong cooling.  
For each of these models we evolve a simulation with
$128^2$ gas and $128^2$ dark matter particles,  
using ASPH to model the hydrodynamics and a
Particle-Mesh technique to follow the gravitational interactions.
We also repeat one model with $256^2$ gas and dark matter
particles, to directly assess any numerical resolution dependencies.

Because both the physics and initial conditions of these models are
scale-free, their physical properties must evolve
self-similarly in time, in accord with analytic scaling laws.  
These rigorous analytic predictions provide a stringent test of our
numerical methods, since numerical parameters like particle mass and
force resolution stay fixed as the system evolves and will therefore
act to break self-similar behavior if they limit the physical accuracy
of the calculation.
We test for self-similar evolution by identifying objects as groups of 
particles within a given overdensity contour at various times, measuring global
properties characteristic of these structures,
and examining the evolution of these quantities over a range
of output times.  In general we find excellent agreement between the
analytically predicted and numerically measured scalings
for the masses, temperatures, Bremsstrahlung luminosities, 
and (to a lesser extent) gas densities over the range of 
expansion factors that we would naively expect to
be accessible based on the simple resolution arguments presented in
\S \ref{Sims.sec}.  
Our most significant result, demonstrated directly in 
Figure~\ref{Mcoolfrac.fig} and elaborated in 
Figures~\ref{fM.fig} and~\ref{TtoTcfM.fig},
is that the fraction of baryonic material
that cools in collapsed objects of specified mass follows the expected
scaling with remarkable accuracy.

One of the impressive successes of hydrodynamic cosmological simulations
with CDM initial conditions and realistic cooling laws is that they
produce dense clumps of cold gas with masses, sizes, and overdensities
comparable to the luminous regions of observed galaxies
(Katz et al. 1992, 1996; Evrard et al.\ 1994; Summers, Davis, \& Evrard 1995;
Frenk et al.\ 1996; Weinberg et al.\ 1997; Pearce 1998).
With reasonable prescriptions for galactic scale star formation, these
objects are converted into dense, tightly bound clumps of stars and
cold gas.  The resulting stellar masses are not sensitive to the 
parameters of these prescriptions, at least within some range,
because the star formation rate is governed mainly by the rate at
which gas cools and condenses onto the central object (Katz et al.\ 1996;
Pearce 1998).  If numerical simulations are to provide accurate
predictions of quantities like the galaxy luminosity function,
star formation rates in high redshift systems, or even the bias between
galaxies and mass, then they must accurately follow the gravitational
collapse, shock heating, and subsequent dissipation and condensation
of baryons into these dense systems.

Our results provide encouraging evidence that cosmological simulation
methods can indeed rise to this challenge.  Specifically, they show
that 2-D calculations that resolve individual systems with a few dozen
to a few hundred particles correctly follow the formation of dissipated
objects.  They also suggest that computing the galaxy baryon mass
function may be an easier problem than computing the cluster X-ray
luminosity function despite the higher densities of the objects
in question and the greater complexity of the physics (with radiative
cooling playing a larger role).  The $\rho^2$ dependence of Bremsstrahlung
emissivity implies that a simulation must resolve an object's central
density in order to compute its X-ray luminosity accurately.
However, thermal instability is a threshold phenomenon, and once a
simulation resolves an object's density at the cooling radius
(where the cooling time equals the dynamical time), it can compute
the dissipated baryon mass with reasonable accuracy.  While it is
clearly important to repeat the self-similar evolution tests with
3-D calculations, we see no reason that the behavior in three dimensions
should be fundamentally different.

For nearly two decades, cosmological N-body simulations have provided
an indispensable tool for predicting the large scale distribution of 
dark matter in different cosmological models.  It appears that the
more ambitious goal of predicting the properties and distribution of 
galaxies with hydrodynamic simulations is now well within reach.

\acknowledgments
This work was supported by NASA grants NAG5-2882 and NAG5-3111.  JMO's work
was partially supported under the auspices of U.S. DOE by LLNL under
contract W-7405-Eng-48.


\end{document}